\documentclass[12pt,draftclsnofoot,onecolumn]{IEEEtran}

\usepackage{amsfonts}
\usepackage{amsmath}
\usepackage{amssymb}
\usepackage{bm}
\usepackage{graphicx}
\usepackage{subfigure}
\usepackage{color, soul}
\usepackage{stfloats}
\usepackage[numbers,sort&compress]{natbib}
\usepackage[amsmath,thmmarks]{ntheorem}
\usepackage{theorem}
\usepackage{mathtools}
\usepackage{threeparttable}

\theoremheaderfont{\sc}\theorembodyfont{\upshape}
\theoremstyle{nonumberplain}
\theoremseparator{}
\theoremsymbol{\rule{1ex}{1ex}}

\begin{document}
\title{One-bit LFMCW Radar: Spectrum Analysis and Target Detection}

\author{Benzhou Jin, Jiang Zhu, \emph{Member, IEEE},, Qihui Wu, \emph{Senior Member, IEEE}, Yuhong Zhang, \emph{Senior Member, IEEE}, and Zhiwei Xu, \emph{Senior Member, IEEE}\thanks{This work is supported in part by the National Natural Science Foundation of China under Grant 61971218, 61901415, 61827801, 61631020, and in part by the Natural Science Foundation of Jiangsu Province under Grant BK20190396. (Corresponding author: Jiang Zhu.)

B. Jin and Q. Wu are with the Key Laboratory of Dynamic Cognitive System of Electromagnetic Spectrum Space, Ministry of Industry and Information Technology, Nanjing University of Aeronautics and Astronautics, Nanjing 211106, China (E-mails: jinbz@nuaa.edu.cn, wuqihui@nuaa.edu.cn).

J. Zhu and Z. Xu are with the Ocean College, Zhejiang University, No.1 Zheda Road, Zhoushan, 316021, China (E-mails: jiangzhu16@zju.edu.cn, xuzw@zju.edu.cn).

Y. Zhang is with the School of Electronic Engineering, Xidian University, Xi’an, Shaanxi 710071, China (E-mail: yuhzhang@xidian.edu.cn).}}

\date{}

\maketitle

\begin{abstract}
One-bit radar, performing signal sampling and quantization by a one-bit ADC, is a promising technology for many civilian applications due to its low-cost and low-power consumptions. In this paper, problems encountered by one-bit LFMCW radar are studied and a two-stage target detection method termed as the dimension-reduced generalized approximate message passing (DR-GAMP) approach is proposed. Firstly, the spectrum of one-bit quantized signals in a scenario with multiple targets is analyzed. It is indicated that high-order harmonics may result in false alarms (FAs) and cannot be neglected. Secondly, based on the spectrum analysis, the DR-GAMP approach is proposed to carry out target detection. Specifically, linear preprocessing methods and target predetection are firstly adopted to perform the dimension reduction, and then, the GAMP algorithm is utilized to suppress high-order harmonics and recover true targets. Finally, numerical simulations are conducted to evaluate the performance of one-bit LFMCW radar under typical parameters. It is shown that compared to the conventional radar applying  linear processing methods, one-bit LFMCW radar has about $1.3$ dB performance gain when the input signal-to-noise ratios (SNRs) of targets are low. In the presence of a strong target, it has about $1.0$ dB performance loss.

\end{abstract}

\section{Introduction}
As radar systems scale up in both bandwidth and the number of antenna elements \cite{Ghelfi, Baransky, Wang, Eldhuset}, high-precision analog-to-digital converters (ADCs) applied in the conventional fully digital radar become a limiting factor \cite{Le,Ali,Li2016,Li22016} for many low-cost and resource-limited applications, such as anti-drone radars \cite{Park} and Google's hand gesture recognition radars \cite{Lien}. Firstly, hundreds or even thousands of high-precision ADCs working  at the Nyquist sampling frequency make the system costly and power-hungry. Secondly, the huge data generated by the antenna array is difficult to transmit, store, access and process.

One-bit radar adopting one-bit ADCs to carry out signal sampling and quantization becomes a promising solution to overcome the above bottlenecks and has recently attracted considerable research interest \cite{Li2016, Li22016, Jacques2018, BjphD, HL2018}. Compared to the conventional radar, one-bit radar has two advantages. Firstly, the one-bit ADC can be implemented inexpensively and energy efficiently through a simple comparator. Secondly, the data rate generated by the antenna array can be largely reduced since the one-bit ADC represents a complex sample using only 2 bits (the real and imgage part of a sample are, respectively, represented using 1 bit). However, because the one-bit ADC is a highly nonlinear device, conventional signal processing methods, e.g., the matched filtering and compressed sensing (CS) based methods, face new challenges. How to perform target detection based on one-bit quantized signals (represented by one-bit signals for simplicity in the following) deserves in-depth study.

\subsection{Related work}

The related signal processing of one-bit radar can be classified into two categories: signal reconstruction based and parameter estimation based methods. From the signal reconstruction point of view, linear processing methods are usually adopted. The spectrum of one-bit signals, sampling frequency and target detection approaches have been studied \cite{SAR1991,SAR1997,Pascazio1998,Li22016,BjphD,HL2018,NoiseRadar}.
As for the parameter estimation method, the goal is to perform target detection and localization directly via the nonlinear processing \cite{Li2016,Jacques2018,Zahabi2019, Zhang2019, Dong2015} such as the CS based algorithm.
\subsubsection{Signal reconstruction based methods}
In \cite{SAR1991}, the spectrum of one-bit  signal is analyzed and synthetic aperture radar (SAR) imaging is considered. It is found that for a low signal-to-noise ratio (SNR) scenario,  the fundamental component of the one-bit  signal is an unbiased replica (apart from a scaling factor) of the original (before the one-bit quantization) noise-free signal, while when SNR is high, after the one-bit quantization the information on the original signal amplitude is totally lost. In addition, the one-bit signal consists of plentiful self-generated high-order harmonics. When the SNR of a scatter is low, i.e., $\rm {SNR} \ll 1$, amplitudes of high-order harmonics decrease quickly with their orders. Conventional linear methods, e.g., the matched filtering, still can  be used for SAR imaging and the performance degradation is small. In \cite{SAR1997,Pascazio1998}, in order to obtain higher quality SAR images, the effects of high-order harmonics are considered. It is indicated that though the amplitudes of high-order harmonics are negligible with  respect to the fundamental component, the spectrum overlapping between the fundamental and high-order harmonics produces a small aliasing effect and leads to some performance loss. From the respect of linear signal processing, oversampling can reduce the spectrum overlapping \cite{Pascazio1998}. It is shown that the performance of the SAR imaging is improved through an $2x$ oversampling. However, oversampling leads to an increase of the involved data dimensions. Moreover, it is impossible to separate high-order harmonics from the fundamental component totally in the frequency domain by oversampling. Linear processing methods are difficult to eliminate the effect of high-order harmonics completely.

In \cite{Li22016}, target detection is studied for one-bit linear frequency modulated continuous wave (LFMCW) radar. In order to suppress high-order harmonics, an unknown dithering scheme is proposed to reduce the received SNRs before one-bit quantization. If the received SNRs are low enough, high-order harmonics can be neglected. Then,  linear processing methods can be applied efficiently to perform target detection.  However, reducing the received SNRs leads to the performance degradation of target detection inevitably.

\subsubsection{Parameter estimation based methods}
In \cite{Li2016}, a $l_1$ minimization based sparse target detection approach is proposed for one-bit LFM pulse radar. Nevertheless, this work only considers target detection in the fast time domain. Subsequently, this approach is extended to the spatial domain in \cite{Jacques2018} and target detection is carried out in fast time domain for each antenna element. In \cite{Zahabi2019}, target recovery is studied for one-bit pulse-Doppler radar in the presence of clutter.  The recovery problem is achieved by a sparse recovery method which leads to an optimization problem. In \cite{Zhang2019}, based on signed measurements of LFMCW radar, range estimation and range-Doppler imaging are studied through the maximum likelihood approach.  In order to reduce the computational complexity, a relaxation-based approach, referred to as the One-bit RELAX algorithm, is proposed. Moreover, the Bayesian information criterion is used to determine the number of scatters. In \cite{Dong2015}, the maximum a posteriori (MAP) approach is used to suppress ghosts caused by high-order harmonics for the one-bit SAR imaging. Results show that the proposed nonlinear recovery method is effective to eliminate high-order harmonics. In \cite{Fettweis1}, it is shown that  compared to the conventional high-precision quantized system, the performance loss caused by the one-bit ADC is more than $2/\pi$ (for the low SNR case, the loss is about $2/\pi$) at the Nyquist sampling rate. Oversampling can reduce the loss for nonlinear processing methods as well \cite{Stein1, Fettweis1, lmaz, Fettweis2}.

Nonlinear processing methods do not concern the spectrum of one-bit signals since they perform parameter estimation directly based on the observations. Nevertheless, when all of the three domains (i.e., the spatial, slow time, and fast time domain) are considered  for  one-bit fully digital radar, the dimension of  observations, as will be seen later, is generally huge, and previous methods are difficult to implement due to the unaffordable computational complexity.

\subsection{Contributions}

In this paper, target detection for  one-bit LFMCW radar is studied. As discussed above, linear processing methods can not suppress high-order harmonics completely. While for nonlinear methods, the huge dimension of the observations leads to difficulties with computation. In this paper, a dimension-reduced generalized approximate message passing (DR-GAMP) approach is proposed for  target detection. The main contributions of this work can be summarized as follows:
\begin{itemize}
  \item The spectrum of the one-bit signal is analyzed. The spectrum analysis results in \cite{SAR1991} are extended by taking cross-generated harmonics into account. In particular, for a  scenario with two targets, the closed-form of the average amplitudes of the cross-generated harmonics are obtained. The  average amplitudes of the 3-order cross-generated harmonics are larger than that of the 3-order self-generated harmonics if the two targets have the same received SNRs, i.e., $\rm SNR_1 = \rm SNR_2$. Based on the spectrum analysis, the effects of one-bit quantization on linear processing methods are investigated. It is shown that the 3-order harmonics can cause false alarms (FAs).
  \item To perform target detection, a two-stage method, termed as the DR-GAMP approach which combines linear and nonlinear processing methods,  is proposed. In the first stage, conventional linear processing methods are implemented to perform the dimension reduction. In the second stage, by exploiting the sparsity of the targets in the searching space, the GAMP algorithm is adopted to suppress the harmonics and recover true targets. It is numerically shown that the DR-GAMP approach benefits significantly from the dimension reduction.
  \item The performance of the proposed method is investigated for both on-gird and off-gird cases through numerical simulations under typical parameters. Results show that the DR-GAMP approach is effective to remove high-order harmonic FAs and perform target detection for one-bit LFMCW radar.
\end{itemize}

The rest of this paper is organized as follows. In Section II, the signal model is introduced. The spectrum of  one-bit  signal is analyzed and its effects on target detection are discussed in Section III. Then the proposed DR-GAMP approach is presented in Section IV. In Section V, substantial numerical experiments are conducted to illustrate the target detection performance of one-bit LFMCW radar. Finally, we conclude the paper in Section VI.
\subsection{Notation}
${\rm csign}(\cdot) = {\rm sign}({\rm Re}(\cdot))+{\rm j}{\rm sign}({\rm Im}(\cdot))$, where ${\rm Re}(\cdot)$ and ${\rm Im}(\cdot)$ are the real and imaginary parts, respectively, and ${\rm sign}(\cdot)$ returns the componentwise sign. ${\rm E}(\cdot)$ denotes the expectation operation. ${\otimes}$ denotes the Kronecker product. ${\mathcal F}(\cdot)$ denotes the Fourier transform. $\|\cdot\|_0$ is the zero pseudo-norm. $\lfloor \cdot \rfloor$ rounds its variable to the nearest integer less than or equal to that variable. $\rm {rect}(\cdot)$ denotes the rectangle function.
\section{Signal model}
\begin{figure}[h!t]
\centering
\includegraphics[width=3.8in]{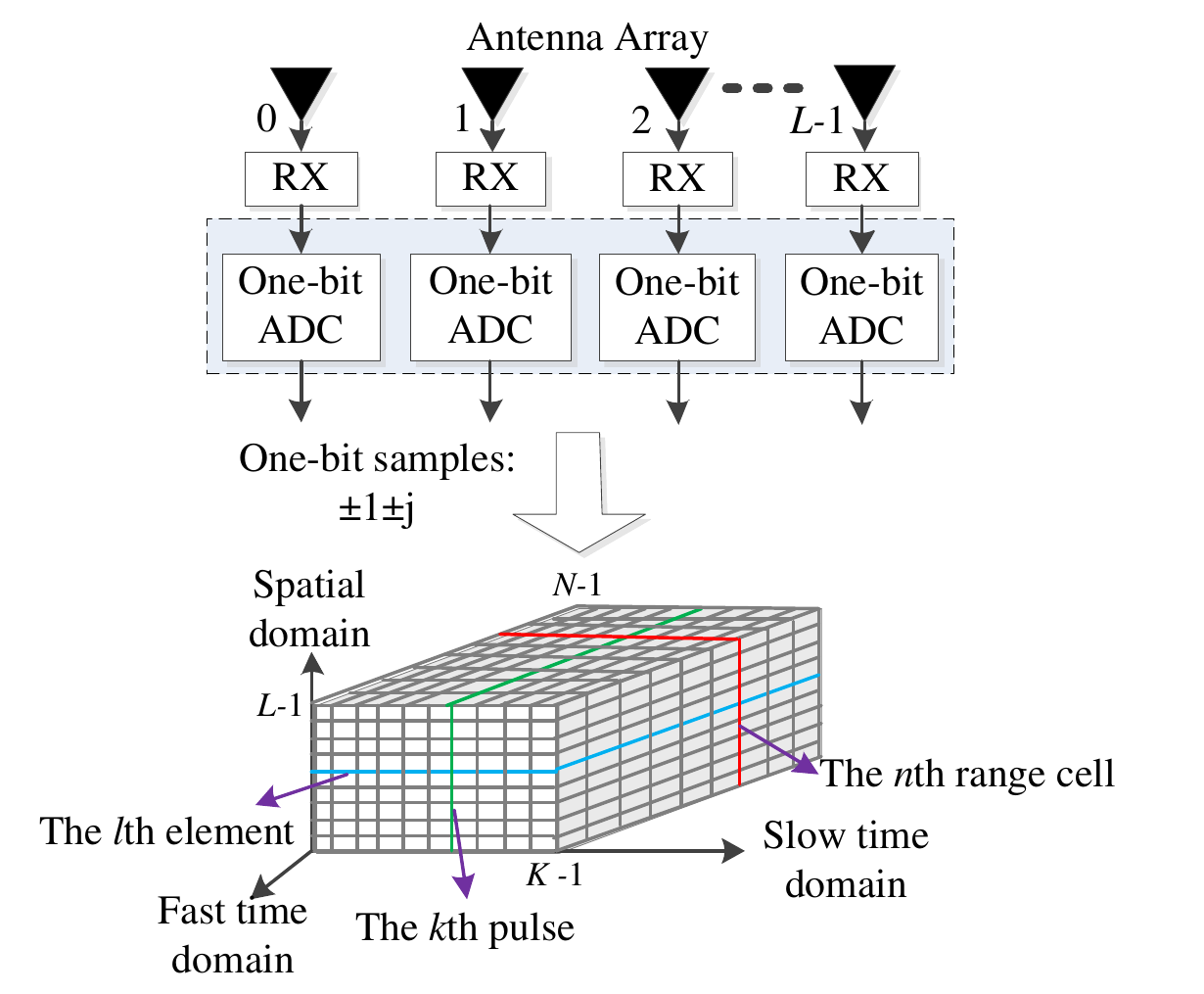}
\caption {The receiver architecture and a general model of data collection for the fully digital one-bit radar. RX denotes the radio frequency channel mainly including the low noise amplifier (LNA), dechirping block and filter. After RX, one-bit ADCs are applied to implement signal sampling and quantization. Then, the baseband data can be collected in a data cube.}
\label{figdatacube_pdf}
\end{figure}

In this paper, the transmitter for one-bit LFMCW radar is the same as the conventional radar and in a coherent pulse interval (CPI), periodic linear frequency modulated pulses are transmitted. Considering a fully digital uniform linear array, the receiver architecture and a general model of data collection is illustrated in Fig. \ref{figdatacube_pdf}. Different from the conventional radar applying high-precision ADCs, one-bit ADCs are adopted.

As shown in Fig. \ref{figdatacube_pdf}, each antenna element includes a separate radio frequency (RF) channel. Before dechirping (also known as the stretch processing), the received signal in a CPI for the $l$th antenna element can be represented as\footnote{Here clutter is neglected. Since the clutter is signal dependent, the one-bit quantization leads to complex coupling among clutter, targets and noise. This paper is to take a first step in providing a  target detection approach for one-bit LFMCW radar without considering clutter.}
\begin{align}
q_l(t)\! \! &= \! \! \sum\limits_{p=1}^P \tilde{\sigma}_{s,p}   {\rm rect} \! \! \left(\frac{t \! - \! kT_I \! - \! {\tau}_p(t)}{T_I}\right) {\rm exp}({\rm -j}2{\pi}f_c(t \! \! - \! \!KT_I \!- \!{\tau}_p(t)) \! \!+ \!\! {\rm j}{\pi}{\mu}(t\!-\!kT_I\!-\!{\tau}_p(t))^2\!+\!{\rm j}2{\pi}f_{sp,p}l) \notag\\& + w_q(t), \quad (0 \leq t \leq KT_I)
\label{receSigModelBefDDC}
\end{align}
where $0\leq l \leq L-1$, $L$ is the number of antenna elements, $f_c$ is the carrier frequency, $P$ is the number of targets, ${\tau}_p(t)$ is the delay of the $p$th target and $\tilde{\sigma}_{s,p}$ is the random, complex voltage. $K$ is the number of pulses in a CPI and $0\leq k\leq K-1$. $T_I$ is the pulse interval and ${\mu}$ is the frequency modulation slope. $w_q(t)$ is the additive white Gaussian noise (AWGN).  $f_{sp,p}=d{\rm sin}{\varphi}_p/{\lambda}$ is the spatial frequency, where ${\varphi}_p$ is the azimuth angle of the $p$th target and $d$ is the inter-element spacing of the antenna array.

Generally, the dechirping is performed in the analog domain to obtain a beat signal as\footnote{The one-bit signal after dechirping is studied in this paper. Applying one-bit ADC directly without dechirping is worth studying and will be left for future work.}
\begin{align}
r_l(t)&= q_l(t){\rm rect} \left(\frac{t-kT_I}{T_I}\right){\rm exp}({\rm j}2\pi f_c(t-KT_I){\rm -j}{\pi}{\mu}(t-kT_I)^2).
\label{receSigModelBefDDC}
\end{align}
Let ${{\tau}}_{\rm max}$ be the maximum target delay and $0<{\tau}_p(t) \leq{{\tau}}_{\rm max}$. The valid observation time interval in the $k$th pulse is $(kT_I+{\tau}_{\rm max},  (k+1)T_I]$. Here, we merely consider the received signal in (\ref{receSigModelBefDDC}) in the valid time duration and (\ref{receSigModelBefDDC}) is further given by
\begin{align}
r_l(t)&=\sum\limits_{p=1}^P \tilde{\sigma}_{s,p} {\rm exp}\left({\rm -j}2{\pi}f_c\tau_p(t)-{\rm j}2{\pi}\mu (t-kT_I)\tau_p(t)+{\rm j}{\pi}\mu \tau_p^2(t) + {\rm j}2{\pi}f_{sp,p}l\right) + w_q(t),
\label{receSigModelBefDDC2}
\end{align}
where ${\tau}_p(t)$ is ${\tau}_p(t)=2(R_{0,p}-v_pt)/c = \tau_{0,p}-2v_p(kT_I+t-kT_I)/c$, $R_{0,p}$ is the initial range of the $p$th target for the current CPI, ${\tau}_{0,p} = 2R_{0,p}/c$ and  $c$ is the speed of light. $kT_I$ and $t-kT_I$ are known as the slow time and fast time, respectively.  Then, (\ref{receSigModelBefDDC2}) can be simplified as
\begin{align}
r_l(t) \! &= \! \! \sum\limits_{p=1}^P {\sigma}_{s,p} {\rm exp} \! \! \left( \! {\rm j}2{\pi}\big(f_{d,p}kT_I \! + \! (f_{r,p} \!+ \!f_{d,p} \! + \! 2\mu v_pt/c)(t \! - \! kT_I) \! + \! f_{sp,p}l \! + \! \frac {1}{2}\mu {\tau}_p^2(t)\big) \! \! \right)\! \! + \! w_q(t),
\label{receSigModelBefDDC3}
\end{align}
where ${\sigma}_{s,p}=\tilde{\sigma}_{s,p}{\rm exp}(-j4\pi R_{0,p}/ {\lambda})$, $f_{d,p}=2v_p/{\lambda}$ is the Doppler shift, $v_p$ denotes the $p$th target velocity. $f_{r,p} = -{\mu}{\tau}_{0,p}$ denotes the beat frequency which represents the range of the $p$th target. In general, $|(f_{d,p}+2\mu v_pt/c)(t-kT_I)|\ll 1$ and $|\mu {\tau}_p^2(t)| \ll 1$. Hence, (\ref{receSigModelBefDDC3}) can be approximated as
\begin{align}
&r_l(t)\approx \sum\limits_{p=1}^P {\sigma}_{s,p} {\rm exp}\left({\rm j}2{\pi}(f_{d,p}kT_I+f_{r,p}(t-kT_I)+f_{sp,p}l)\right)\!+ w(t),
\label{receSigModelBefDDCApp}
\end{align}
The signal bandwidth of target echoes in the fast time domain is no more than $| \mu \tau _{\rm max}|$, i.e., $f_{r,p}\in[-B_r,0]$, where $B_r = | \mu \tau _{\rm max}|$. Hence, the received signal bandwidth is limited to $B_r$ by a bandpass filter. Meanwhile, the bandwidth of noise is reduced to $B_r$ as well.  After one-bit ADC, the received data in one CPI can be represented by a $K{\times}L{\times}N$ data cube, as shown in Fig. \ref{figdatacube_pdf}. The element $r(k,l,n)$ of the data cube can be modeled as
\begin{align}
r(k,l,n)= {\rm csign} \left(\sum\limits_{p=1}^P{\sigma}_{s,p}{\rm exp}({\rm j}2{\pi}(f_{r,p}nT_s+f_{d,p}kT_I+f_{sp,p}l))+w(k,l,n)\right),\label{receSigModel}
\end{align}
where $0\leq n\leq N-1$ and $N$ is the number of samples in the fast time domain within the valid observation time duration $T$. $T_s = 1/f_s$ is the sampling interval.

For the conventional radar applying high-precision ADCs, linear signal processing methods are generally implemented independently in the three domains. In contrast, as will be seen, the nonlinear method is applied jointly to  process the received data. In order to satisfy the later requirement of the nonlinear processing, we subsequently represent the received signal in matrix forms.

The slow time domain is firstly considered. The Doppler shift interval $[0, {\rm PRF}]$ is discretized into $M_d$ grid points, where ${\rm PRF} = 1/T_I$ denotes the pulse repetition frequency (PRF). In order to establish the signal model, we assume  the Doppler shifts of targets lie on the grid. In general, $M_d$  satisfies $({\rm PRF}/M_d)\! \leq \!{\Delta}f_d$, i.e., $M_d{\geq}K$, where ${\Delta}f_d \!= \! 1/(KT_I)$ denotes the Doppler frequency resolution. Construct a dictionary matrix ${\mathbf A_d}\in{\mathbb C}^{{K\times}M_d}$ and let ${\mathbf a}_d(f_{d,m_d}) = \left[1, {\rm exp}({\rm j}2{\pi}f_{d,m_d}T_I),...,{\rm exp}\left({\rm j}2{\pi}(K-1)f_{d,m_d}T_I\right)\right]^{\rm T}$ be its $m_d$th column, where $0\leq m_d\leq M_d-1$ denotes the index of the grid point in the slow time domain and $f_{d,m_d} = (m_d{\rm PRF})/M_d$. Then, for the $l$th antenna element and $n$th range cell, the received signal can be  expressed as
\begin{align}\label{MTSlowTimeSigModel}
{\mathbf r}_d(l,n)={\rm csign} \left({\mathbf A}_d{\mathbf x}_d(l,n)+{\mathbf w}_d(l,n)\right),
\end{align}
where ${\mathbf x}_d(l,n)=[x_{d,0}(l,n),...,x_{d,m_d}(l,n),...,x_{d,M_d-1}(l,n)]^{\rm T}{\in}{\mathbb C}^{M_d{\times}1}$ and the number of nonzero element of ${\mathbf x}_d(l,n)$ satisfies $\|{\mathbf x}_d(l,n)\|_0\leq P$. The inequality comes from the fact that targets with different spatial or beat frequencies may share the same Doppler shift, otherwise equality holds. In practice, targets do not lie on the grid exactly. Nevertheless, when the dictionary is densely enough, (\ref{MTSlowTimeSigModel}) is valid and ${\mathbf x}_d(l,n)$ is approximately sparse \cite{mismatch}.

Similarly, the received signal in the spatial and fast time domain can be described by
\begin{align}\label{MTSpaSigModel}
{\mathbf r}_{sp}(k,n)={\rm csign} \left({\mathbf A}_{sp}{\mathbf x}_{sp}(k,n)+{\mathbf w}_{sp}(k,n)\right),
\end{align}
and
\begin{align}\label{MTFastTimeSigModel}
{\mathbf r}_r(k,l)={\rm csign} \left({\mathbf A}_r{\mathbf x}_r(k,l)+{\mathbf w}_r(k,l)\right),
\end{align}
where ${\mathbf x}_{sp}(k,n)=[x_{sp,0}(k,n),...,x_{sp,m_{sp}}(k,n),...,x_{sp,M_{sp}-1}(k,n)]^{\rm T}{\in}{\mathbb C}^{M_{sp}{\times}1}$ and $M_{sp}$ denotes the number of grid points in spatial domain. ${\mathbf x}_r(k,l)\!\!=\!\![x_{r,0}(k,l), ...,x_{r,m_r}(k,l),...,x_{r,M_r-1}(k,l)]^{\rm T} $ $ {\in}{\mathbb C}^{M_r{\times}1}$ and $M_{r}$ denotes the number of grid points in fast time domain. $m_{sp}$ and $m_{r}$ are the grid point indices in the spatial and fast time domains, respectively. Generally, $M_{sp}{\geq}L$ and $M_r{\geq}N$. We assume spatial frequencies and beat frequencies of targets lie on the grid as well. Similar to the case in the slow time domain, the number of nonzero elements of  ${\mathbf x}_{sp}(k,n)$ and ${\mathbf x}_r(k,l)$ satisfy $\|{\mathbf x}_{sp}(k,n)\|_0\leq P$ and $\|{\mathbf x}_r(k,l)\|_0\leq P$, respectively. At the $m_{sp}$th grid point in the spatial domain, the spatial frequency is $f_{{sp},m_{sp}} = m_{sp}/M_{sp}$. For the fast time domain, the beat frequency at $m_r$th grid point is $f_{r,m_r} = -2{\mu}R_{m_r}/c$, where $R_{m_r}$ is the target range corresponding to the $m_r$th grid point. ${\mathbf A}_{sp}{\in}{\mathbb C}^{L{\times}M_{sp}}$ is the dictionary matrix in the spatial domain and its $m_{sp}$th column is ${\mathbf a}_{sp}(f_{sp,m_{sp}} \!)  \!\!= \!\! \left[1, {\rm exp}({\rm j}2{\pi}f_{sp,m_{sp}}),...,{\rm exp} \!\!\left({\rm j}2{\pi}(L \!\!- \!\!1)f_{sp,m_{sp}} \!\right)\right]^{\rm T}$. ${\mathbf A}_r$ is the dictionary matrix in the fast time domain and its $m_r$th column is ${\mathbf a}_r(f_{r,m_r\!}) \!\! = \!\! \left[1, {\rm exp}({\rm j}2{\pi}f_{r,m_r\!}),...,{\rm exp}\left({\rm j}2{\pi}(N\!\!-\!\!1)f_{r,m_r \!}\right)\right]^{\rm T}$.

Combining (\ref{MTSlowTimeSigModel}), (\ref{MTSpaSigModel}) and (\ref{MTFastTimeSigModel}), the received data cube can be reduced to a $KLN{\times}1$ column vector which is given by
\begin{align}\label{receSigMatrForm}
{\mathbf r} = {\rm csign} \left({\mathbf A}{\mathbf x}+{\mathbf w}\right),
\end{align}
${\mathbf x} \!= \! \! [x(0),...,x(m_dM_{sp}M_r \!+ \!m_{sp}M_r\!+m_r),...,x(M_dM_{sp}M_r-1)]^{\rm T}{\in}{\mathbb C}^{M_dM_{sp}M_r{\times}1}$ and $x(m_d(M_{sp}M_r)+m_{sp}M_r+m_r)$ denotes the complex amplitude of the target whose Doppler shift, spatial frequency and beat frequency are $f_{d,m_d}$, $f_{sp,m_{sp}}$ and $f_{r,m_r}$, respectively, ${\mathbf A}={\mathbf A}_d{\otimes}{\mathbf A}_{sp}{\otimes}{\mathbf A}_r{\in}{\mathbb C}^{LKN{\times}M_dM_{sp}M_r}$. In the three signal domains, targets are generally sparse and typically, the number of nonzero elements of ${\mathbf x}$ satisfies $\|{\mathbf x}\|_0=P \ll M_dM_{sp}M_r$.

Because the one-bit ADC is a highly nonlinear device, conventional linear processing methods, as shown in the ensuing Section \ref{spectrum}, may be ineffective because of harmonics. While for nonlinear reconstruction methods, they are generally impossible to be implemented because of the large dimension of ${\mathbf A}$ in (\ref{receSigMatrForm}). For example, assuming the vector ${\mathbf r} $ and ${\mathbf x}$ both have the dimension of $10^6$, the matrix ${\mathbf A}$ contains $10^{12}$ entries which leads to difficulties with computation and memory.
\section{Spectrum analysis of the one-bit quantized signal}\label{spectrum}
For target detection in one-bit radar, a natural question is whether linear processing methods, e.g., Fast Fourier transform (FFT),  are still effective or not. To answer this question, the signal spectrum of the one-bit signal is investigated in a  scenario with multiple targets. An approximation of the one-bit signal  is obtained in the frequency domain based on which, effects of one-bit quantization on target detection are discussed.
\subsection{Harmonics analysis}


In this subsection, the real part of the received complex one-bit signal is firstly analyzed and then, results are extended to the complex one-bit signal. As shown in (\ref{receSigModel}), the received signal models are similar in the three domains.  Without loss of generality, we take the signal in fast time domain  for example to analyze. In the continuous form, with both indices of the slow time and spatial domain fixed, the received signal of one pulse in the fast time domain can be abstracted as
\begin{align}
v_q(t)={\rm rect}\left(\frac {t}{T}\right){\rm sign}\left( \sum\limits_{p=1}^PA_p\cos(\omega_pt+\Phi_p)+w(t) \right)
\end{align}
Since the term ${\rm rect}\left(\frac {t}{T}\right)$ has no effect on harmonic characteristics and its effects on spectrum of $v_q(t)$ is known, we ignore ${\rm rect}\left(\frac {t}{T}\right)$ in the following for simplicity. Let $u(t)=\sum\limits_{p=1}^PA_p\cos(\omega_pt+\Phi_p)$. The one-bit signal $v_q(t)$ is given by \cite{SAR1991}
\begin{align}\label{decom}
v_q(t)&={\rm sign}(u+w)=-\frac{\rm j}{\pi}\int_{-\infty}^{\infty}\frac{{\rm exp}({\rm j}(u+w)\xi)}{\xi}{\rm d}\xi\notag\\
&=-\frac{\rm j}{\pi}\int_{-\infty}^{\infty}\frac{{\rm exp}({\rm j}w\xi)}{\xi}\left(\prod\limits_{p=1}^P{\rm e}^{{\rm j} A_p \cos(\omega_pt+\Phi_p)\xi}\right){\rm d}\xi\notag\\
&\overset{(a)}=-\frac{\rm j}{\pi}\int_{-\infty}^{\infty}\frac{{\rm exp}({\rm j}w\xi)}{\xi}\prod\limits_{p=1}^P   \left(\sum\limits_{m_p=0}^{\infty}\epsilon_{m_p}{\rm j}^{m_p}J_{m_p}(A_p\xi)\cos({m_p\omega_pt+m_p\Phi_p})\right){\rm d}\xi,
\end{align}
where $J_{m_p}(\cdot)$ is the Bessel function of the first kind and equality (a) follows from \cite[pp. 361]{Hyp}. $\epsilon_0=1$ and $\epsilon_m=2$ for $m\geq 1$, where $m=m_1+...+m_P$ denotes the term order. From equation (\ref{decom}), it is shown that $v_q(t)$ not only preserves their original frequencies ( also known as fundamental harmonics and $m = 1$), but also includes new frequencies (high-order harmonics). These high-order harmonics contain  both cross-generated terms from  different sinusoidals and self-generated terms from their original sinusoidals.  Amplitudes of both original and new frequencies in $v_q(t)$ are related to the noise $w(t)$, i.e., these amplitudes are all random variables. As will be seen in Section IV, the one-bit signal is first processed using  FFT. Clearly,  FFT can be regarded as an averaging process over the noise. Since the number of observation points is very large in radar, similar to \cite{SAR1991}, amplitudes of both original and new frequencies can be represented approximately by their average values after FFT. The average value of $v_q(t)$ is
\begin{align}\label{decom_ave}
{\rm E}(v_q(t))&=-\frac{\rm j}{\pi}\int_{-\infty}^{\infty}\frac{1}{\sqrt{2\pi\sigma_w^2}}{\rm e}^{-\frac{w^2}{2\sigma_w^2}}\int_{-\infty}^{\infty}\frac{{\rm e}^{{\rm j }w\xi}}{\xi}\prod\limits_{p=1}^P   \left(\sum\limits_{m_p=0}^{\infty}\epsilon_{m_p}{\rm j}^{m_p}J_{m_p}(A_p\xi)\cos({m_p\omega_pt+m_p\Phi_p})\right){\rm d}w{\rm d}\xi\notag\\
&=-\frac{\rm j}{\pi}\int_{-\infty}^{\infty}\frac{{\rm e}^{-\frac{\sigma_w^2\xi^2}{2}}}{\xi}\prod\limits_{p=1}^P   \left(\sum\limits_{m_p=0}^{\infty}\epsilon_{m_p}{\rm j}^{m_p}J_{m_p}(A_p\xi)\cos({m_p\omega_pt+m_p\Phi_p})\right){\rm d}\xi ,
\end{align}
where the expectation ${\rm E}(\cdot)$ is taken with respect to the noise $w$ and $\sigma_w^2$ is the noise power.  For an even $m$,  both the average amplitudes of self-generated and cross-generated harmonics are equal to zero (refer to Appendix \ref{SA}). For an odd $m$, calculating the average amplitudes of the harmonics is rather clumsy and not necessary for a general $P$. In \cite{SAR1991}, the spectrum under $P=1$ has been analyzed. Here, we consider the $P=2$ case to reveal the characteristics of both self-generated and cross-generated harmonics. Detailed derivations are moved to Appendix \ref{SA} and here, we summarize the main results and provide interesting insights. For completeness, we also outline the results of the $P = 1$  case.

For $P=1$,  ${\rm E}(v_q(t))$ is given by \cite{SAR1991}
\begin{align}\label{ave_spe}
{\rm E}(v_q(t)) = \!\!\!\!\! \sum\limits_{m=1, m~{\rm odd}}^{\infty} \!\!\!\!\! c_m\cos({m\omega t+m\Phi}),
\end{align}
where
\begin{align}\label{sigTargAmpCoff}
c_m=-{\rm j}^{m+1}\sqrt{\frac{2}{{\pi}}}\alpha_m \left(\frac{A_{1}}{\sigma_w}\right)^mF\left(\frac{m}{2};m+1;-\frac{A_{1}^2}{2\sigma_w^2}\right ),
\end{align}
$F(\frac{m}{2};m+1;-\frac{A_{1}^2}{2\sigma_w^2})$ is the hypergeometric function \cite[pp. 504]{Hyp},
\begin{align}
\alpha_m=\frac{1}{(\frac{m-1}{2})!m}2^{-\frac{3(m-1)}{2}}, \quad m ~{\rm is~ odd}.
\end{align}
For example, $\alpha_1=1$, $\alpha_3=1/24$, $\alpha_5=1/640$.
When the received SNR satisfies\footnote{When $\rm{SNR}>>1$, the amplitude information is totally lost \cite{SAR1991}. That is, $c_m$ in (\ref{ave_spe}) is not related with $A_1$. In this paper, we do not consider the high SNR case and it is usually practical for many applications.} ${\rm SNR} = {A_{1}^2}/2{\sigma_w^2}\ll1$, $F(\cdot,\cdot,\cdot)\approx 1$, for a nonnegative integer value $m$ we have
\begin{align}\label{harAmp}
|c_m|\approx \sqrt{\frac{2}{{\pi}}}\alpha_m \left(\frac{A_{1}}{\sigma_w}\right )^m.
\end{align}
According to (\ref{harAmp}), we have
\begin{align}\label{SNRDegr}
20\log \frac{|c_m|}{|c_1|} \approx 20(m-1)\log \frac{A_{1}}{\sigma_w}+20\log \frac{\alpha_m}{\alpha_1}.
\end{align}
(\ref{SNRDegr}) shows that the amplitudes of high-order harmonics decrease rapidly with the harmonic order $m$. Nevertheless, whether high-order harmonics can be omitted or not is determined by the following three factors: the received ${\rm SNR} $, the order $m$ and the digital integration gain of radar. As will be seen later, the 3-order harmonic can not be omitted for many applications. For $m>5$, the amplitudes of $m$-order harmonics are really small and can be omitted. For $m=5$, 5-order harmonics are merely needed to be considered in some scenarios with strong scatters.

For $P=2$, ${\rm E}(v_q(t))$ can be viewed as the superposition of the self-generated and cross-generated harmonics (Note that for $P=1$, ${\rm E}(v_q(t))$ only contains self-generated harmonics). The general expression of the average  amplitudes are given in (\ref{m-order-res_appen}) and (\ref{cross-order-res_appen}) (refer to Appendix \ref{SA}). Here we consider the special that $A_1=A_2$ and reveal characteristics about the self-generated and cross-generated harmonics.

For the $m$-order self-generated harmonics, without loss of generality, we only consider the coefficient $c_{m,0}$ from the first target signal $A_1\cos(\omega_1t+\Phi_1)$. From (\ref{m-order-res_appen}) in Appendix \ref{SA}, the coefficient is given by
\begin{align}\label{m-order-res_furth1}
c_{m,0}=-{\rm j}^{m+1}\sqrt{\frac{2}{{\pi}}}\frac{1}{(\frac{m-1}{2})!m}2^{-\frac{3(m-1)}{2}} ~_3F_3\left(\frac{m+1}{2},\frac{m}{2}+1,\frac{m}{2};1,m+1,m+1;-\frac{2A_2^2}{\sigma_w^2}\right).
\end{align}
When ${\rm SNR}\ll1$, $F(\cdot,\cdot,\cdot)\approx 1$ and we have
\begin{align}\label{m-order-res_furth2}
|c_{m,0}|\approx \sqrt{\frac{2}{{\pi}}}\alpha_m \left(\frac{A_1}{\sigma_w}\right)^m.
\end{align}
The above result is the same as the $P=1$ case as shown in (\ref{harAmp}) .

For the $m$-order cross-generated term $c_{m_1,m_2}\cos(m_2\omega_2 t+m_2\Phi_2)\cos(m_1\omega_1 t+m_1\Phi_1)\notag$, if $A_1=A_2$, the coefficient $c_{m_1,m_2}$ is calculated by (refer to (\ref{cross-order-res_appen}) in Appendix \ref{SA})
\begin{align}\label{cross-order-res_furth1}
c_{m_1,m_2}
=-{\rm j}^{m+1}\sqrt{\frac{2}{{\pi}}}2^{-m+2}\frac{(m-2)!!}{m_1!m_2!} \left(\frac{A_{1}}{\sigma_w}\right)^m ~_3F_3\left(\frac{m+1}{2},\frac{m}{2}+1,\frac{m}{2};m_2+1,m_1+1,m+1;-\frac{2A_2^2}{\sigma_w^2}\right).
\end{align}
where $(m-2)!!=1\times 3\times 5\cdots \times(m-2)$ and $m=m_1+m_2$.

When ${\rm SNR}\ll1$, we have
\begin{align}\label{cross_res}
c_{m_1,m_2}\approx -{\rm j}^{m+1}\sqrt{\frac{2}{{\pi}}}\alpha_{m_1,m_2}^{'} \left(\frac{A_{1}}{\sigma_w}\right)^m,
\end{align}
where
\begin{align}
\alpha_{m_1,m_2}^{'}=2^{-m+2}\frac{(m-2)!!}{m_1!m_2!}.
\end{align}

Comparing  the average amplitude of the 3-order self-generated harmonic with that of the 3-order cross-generated harmonic, we have
\begin{align} \label{harmComp}
20\log \frac{\alpha_{1,2}^{'}/2}{c_{3,0}}=20\log 3\approx 9.5~ {\rm dB},
\end{align}
which demonstrates that for the case $A_1=A_2$, the average energy of the 3-order cross-generated harmonic is $9.5$ dB higher than that of the 3-order self-generated harmonics. As a result, the 3-order cross-generated harmonic may have a stronger effect on target detection in practice.

Further, we consider a complex signal. For $P = 1$, the complex signal after one-bit quantization is
\begin{align}\label{ComSignal}
u_C(t)={\rm csign} \left( A_1{\rm e}^{{\rm j}(\omega t+\Phi)} \right)= {\rm sign} \left(A_1\cos{(\omega t+\Phi)} \right) + {\rm j} \left({\rm sign} A_1\cos{(\omega t+\Phi-\pi/2)})\right).
\end{align}
Its spectrum can be obtained through analyzing the real and imaginary parts, respectively. Compared with the real signal, we need to pay attention to the frequency localizations of high-order harmonics. Taking the 3-order harmonic for example, according to (\ref{ave_spe}) and (\ref{ComSignal}), the 3-order harmonic is given by
\begin{align} \label{ComSignal_3_Order}
c_{3}\cos{(3\omega t+3\Phi)}+{\rm j}c_{3}\cos{(3\omega t+3\Phi-3\pi/2)}=c_{3}{\rm e}^{{\rm -j}(3\omega t+3\Phi)}.
\end{align}
(\ref{ComSignal_3_Order}) means that the frequency of the self-generated 3-harmonic is $-3\omega$. Furthermore, for $P=2$, it can be calculated that  $3$-order harmonics contain $6$ components and their corresponding frequencies are $-3w_1$, $-3w_2$, $2w_1-w_2$, $2w_2-w_1$, $-2w_1-w_2$, and $-2w_2-w_1$, respectively. For a general $P$,  we can similarly obtain the frequency localizations of  $m$-order harmonics. For the complex signal, the average amplitudes of harmonics can be easily calculated based on the results of the real signal and we do not discuss again.

Note that for fundamental components, its bandwidth is still $B_r$, where $B_r$ denotes the bandwidth of the received signal before the one-bit quantization. Nevertheless, the bandwidth of the one-bit signal is larger than $B_r$ because of the existence of high-order harmonics.
Taking the 3-order harmonic for example, as shown in (\ref{ComSignal_3_Order}), its  bandwidth  is $3B_r$.

\subsection{Approximation of the one-bit signal in the frequency domain}
We still take the signal in fast time domain for example and omit ${\rm rect}\left(\frac {t}{T}\right)$. The received one-bit complex signal in continuous form can be described as
\begin{align}\label{fastSig}
v_{qC}(t)={\rm csign } \left(\sum\limits_{p=1}^PA_p{\rm exp}({\rm j}(w_pt+\Phi_p))+w(t)\right).
\end{align}
For the $p$th target, its received SNR before one-bit quantization is defined as ${\rm SNR}_p =   {|A_p|}^2 \!/ \!(2{\sigma}_w^2)$, where ${\sigma}_w^2$ is the noise power of the real part or image part of $w(t)$.

For many applications, since $m$-order harmonics ($m>5$) can be generally neglected, the average value of ${\mathcal F}(v_{qC}(t))$ can be approximated as
\begin{align}\label{cons_harmo}
{\rm E}\left({\mathcal F}(v_{qC}(t))\right )\approx \sum\limits_{q=1}^QA_q\delta{(f-f_q)},
\end{align}
where $Q$ is the number of main components in (\ref{fastSig}) including the fundamental and $3/5$-order harmonics.

On the other hand, based on the previous results in (\ref{decom_ave}), the main components of $v_{qC}(t)$ can be represented approximately by their average values after the Fourier transform. Taking the noise into account, we have
\begin{align}\label{defwf}
{\mathcal F}(v_{qC}(t))\approx {\rm E}\left({\mathcal F}(v_{qC}(t))\right )+w(f).
\end{align}
Here, $w(f)$ is the noise in the frequency domain and we assume  $w(f)$ is the AWGN . Such an assumption will be validated through numerical experiments in Section V. According to the approximations in (\ref{cons_harmo}) and (\ref{defwf}), the one-bit signal after performing the Fourier transform can be approximated as
\begin{align}\label{Gau_ass}
{\mathcal F}(v_{qC}(t))\approx \sum\limits_{q=1}^QA_q\delta{(f-f_q)}+w(f).
\end{align}

The above assumption shows the following two important points:
\begin{itemize}
   \item When the signal (\ref{fastSig}) is sampled, the conventional FFT processing method is still effective for improving the SNRs of both fundamental and high-order harmonics.
  \item Based on the FFT results, we can design a CFAR detector to detect both fundamental and high-order harmonics.  The fundamental components correspond to true targets, while high-order harmonics detected are called as ghosts or FAs that need to be suppressed.
\end{itemize}
\section{A Two-stage DR-GAMP Target Detection Approach}\label{DRGAMP}
\begin{figure}[h!t]
\centering
\includegraphics[width=6.8in]{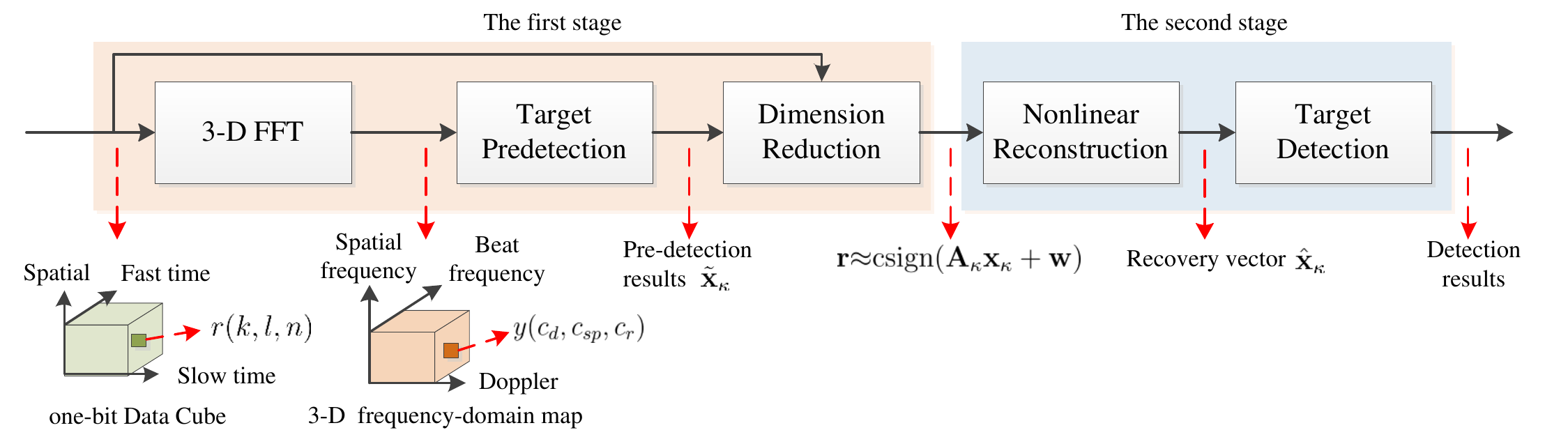}
\caption{ A two-stage DR-GAMP target detection approach. In the first stage, the 3-D FFT and a pre-detection procedure are applied to perform dimension reduction. In the second stage, the GAMP algorithm is adopted firstly  to suppress FAs and then, target detection is carried out based on reconstruction results of the GAMP algorithm.}
\label{TwoStageScheme0418}
\end{figure}

Section \ref{spectrum} shows that for the one-bit signal in (\ref{receSigModel}), the conventional linear approach is effective for improving target SNRs. However, high-order harmonics can lead to FAs and they are difficult to be suppressed. To overcome the drawback, the GAMP algorithm which takes the quantization effects into account is proposed. Nevertheless, because the number of grid points in the target space is very large, the GAMP algorithm can not be performed directly. As a result, a two-stage target detection method, termed as the DR-GAMP  approach,  is proposed, as shown in Fig. \ref{TwoStageScheme0418}. At the first stage, the efficient linear processing is applied in the three domains separately to improve target SNRs and then a pre-detection procedure is carried out to perform dimension reduction. At the second stage, the GAMP algorithm is implemented to recover true targets and suppress harmonics simultaneously.

\subsection{Dimension reduction}
The 3-D FFT is firstly performed over the slow time, spatial, and fast time domain of the received one-bit data cube, respectively. By the 3-D FFT, the SNRs of the fundamental frequencies and high-order harmonics of the received one-bit signal can be all integrated coherently. Assuming the FFT pionts in the slow time, spatial, and fast time domain are $C_d$, $C_{sp}$ and $C_r$, the element of the output 3-D frequency-domain map  can be represented as
\begin{align}\label{3DFFToutPut}
y(c_d,c_{sp},c_r) = {\mathbf a}^{\rm H}(c_d,c_{sp},c_r){\mathbf r}, \quad 0{\leq}c_d{\leq}C_d-1, 0{\leq}c_{sp}{\leq}C_{sp}-1,0{\leq}c_r{\leq}C_r-1.
\end{align}
where ${\mathbf r}$ is the received data in the vector form, as shown in ({\ref{receSigMatrForm}}). ${\mathbf a}(c_d,c_{sp},c_r)$ is the steering vector of the current cell in frequency domain and
\begin{align}\label{3DsteerVector}
{\mathbf a}(c_d,c_{sp},c_r)={\mathbf a}_d(f_{d,c_d}){\otimes}{\mathbf a}_{sp}(f_{sp,c_{sp}}){\otimes}{\mathbf a}_r(f_{r,c_r}),
\end{align}
where $f_{d,c_d}$, $f_{sp,c_{sp}}$ and $f_{r,c_r}$ denotes the Doppler, spatial and beat frequency, respectively, and $f_{d,c_d} = (c_d{\rm PRF})/C_d$, $f_{sp,c_{sp}}=c_{sp}/C_{sp}$ and $f_{r,c_r}=(c_rf_s)/C_r$.

Subsequently, target predetection is implemented for the 3-D frequency-domain map. In this paper, we apply the order statistic constant false alarm rate (OS CFAR) detector\footnote{Here, the detector is CFAR since the noise of FFT output is AWGN.} \cite[pp. 371]{Richards}. The threshold $\gamma_1$ of the OS CFAR detector is
\begin{align}\label{gamma1set}
{\gamma}_{1} = \alpha_{OS}x_\eta,
\end{align}
where $x_\eta$ is the $\eta$th order statistic of reference cells, $\alpha_{OS}$ is a scale factor.  Target predetection is sequentially performed over the Doppler, spatial and beat frequency domain, respectively. Let $I_{pd}$ denote the number of pre-detection targets (PTs). PTs consist of three parts, i.e., true targets, FAs caused by the  high-order harmonics and FAs caused by noise. $\gamma_1$ is chosen based on the noise FA rate $P_{{\rm FA},w}$. Under the assumption that $w(f)$ is the AWGN in (\ref{Gau_ass}), $\gamma_1$ can be easily set based on given $P_{{\rm FA},w}$. Typically, we can set a very small $P_{{\rm FA},w}$, e.g., ranging from $P_{{\rm FA},w}=10^{-3}\sim10^{-6}$. On the other hand, for most practical scenarios, true targets and their $m$-order harmonics (consider $m\leq 5$ case) are sparse in the 3-D frequency data cube as well. That is, the FA rate $P_{{\rm FA},h}$ caused by high-order harmonics satisfies $P_{{\rm FA},h} \ll 1$ as well. Then, we have $I_{pd} \ll C_dC_{sp}C_r$. Let $\tilde{\mathbf x}_{\kappa} = [\tilde{x}_{\kappa}(0),..., \tilde{x}_{\kappa}(i_{pd}), ...\tilde{x}_{\kappa}(I_{pd}-1)]^{\rm T} \in {\mathbb C} ^{I_{pd} \times 1}$  be the pre-detection vector, where $\tilde{{{x}}}_{\kappa}(i_{pd})$ denotes the complex amplitude of the $i_{pd}$th PT whose corresponding frequency cell indices are $c_{d,i_{pd}}$, $c_{sp,i_{pd}}$ and $c_{r,i_{pd}}$, respectively. Note that $0\leq c_{d,i_{pd}}\leq C_d-1$, $0\leq c_{sp,i_{pd}}\leq C_{sp}$ and $0\leq c_{r,i_{pd}} \leq C_r$.

In this paper, we let $M_d = C_d$, $M_{sp} = C_{sp}$ and $M_r = C_r$. Meanwhile, $m_d$, $m_{sp}$ and $m_r$ satisfy $m_d = c_d$, $m_{sp} = c_{sp}$, $m_r = c_r$, respectively. Considering the definition of $m_d$, $m_{sp}$ and $m_r$, the frequency cell indices of $c_d$, $c_{sp}$ and $c_r$ can be regarded as the indices of grid points in the slow time, spatial, and fast time domain as well. The number of grid points in the three domains can be controlled by adjusting the FFT points. Here, we introduce a overgriding factor
\begin{align}\label{defovergriding}
r_a \triangleq M_d/K = M_{sp}/L = M_r/N \geq 1.
\end{align}
Obviously, a higher value of $r_a$ indicates larger number of grid points in the three domains. As will be seen in Section V, when targets are off-grid, in order to suppress high-order harmonics, $r_a$ should be increased.

Based on the above discussion, the number of pre-detection targets  satisfies $I_{pd} \ll C_dC_{sp}C_r = M_dM_{sp}M_r$.  Since pre-detection results reveal the grid point indices of the main components in the vector ${\mathbf r}$, we can obtain a DR model of (\ref{receSigMatrForm}) given by\footnote{Note that for the off-grid case, though biases exist between the true frequencies  and the predetected grid points, we can use (\ref{OnebitReceSigApprox}) to represent $\mathbf {r}$ as well. Performance of the proposed target detection approach for the off-grid case will be investigated in Section \ref{NS}.}
\begin{align}\label{OnebitReceSigApprox}
{\mathbf r} {\approx} {\rm csign}({\mathbf A}_{\kappa}{\mathbf x}_{{\kappa}}+{\mathbf w}).
\end{align}
where ${\mathbf x}_{{\kappa}}=[x_{{\kappa}}(0),..., x_{{\kappa}}(i_{pd}),...,x_{{\kappa}}(I_{pd}-1)]{\in}{\mathbb C}^{I_{pd}{\times}1}$.  For the $i_{pd}$th entry, if $x_{{\kappa}}(i_{pd})=0$, target is absent. Otherwise, $x_{{\kappa}}(i_{pd})$ denotes the true complex amplitude of the target whose Doppler, spatial and beat frequencies are $f_{d,i_{pd}} = (m_{d,i_{pd}}{\rm PRF})/M_d$, $f_{sp,i_{pd}} = m_{sp,i_{pd}}/M_{sp}$ and $f_{r,i_{pd}} = (m_{r,i_{pd}}f_s)/M_r$, respectively.  ${\mathbf A}_{\kappa}=[{\mathbf a}_{\kappa}(0),...,{\mathbf a}_{\kappa}(i_{pd}),...,{\mathbf a}_{\kappa}(I_{pd}-1)]{\in}{\mathbb C}^{KLN{\times}I_{pd}}$ is the DR observation matrix where ${\mathbf a}_{\kappa}(i_{pd})={\mathbf a}_d(f_{d,i_{pd}}){\otimes}{\mathbf a}_{sp}(f_{sp,i_{pd}}){\otimes}{\mathbf a}_r(f_{r,i_{pd}})$.
Based on the known ${\mathbf A}_{{\kappa}}$ and ${\mathbf r}$, we can try to reconstruct the vector ${{\mathbf x}}_{\kappa}$ instead of ${{\mathbf x}}$ in (\ref{receSigMatrForm}).

\subsection{The GAMP based target reconstruction and detection}
Mathematically, the goal of the target reconstruction problem is to find a vector ${\mathbf x}_{{\kappa}}$ consistent with the observation model in (\ref{OnebitReceSigApprox}). Here, we recover ${\mathbf x}_{{\kappa}}$ via Bayesian methods.

The  target reconstruction can be abstracted as recovery problem in the generalized linear model (GLM), where the vector ${\mathbf x}_{\kappa}$ follows i.i.d. prior $p({\mathbf x}_{\kappa})=\prod\limits_{i=1}^{I_{pd}}p({x}_{\kappa_i})$ and undergoes a linear transform ${\mathbf z}={\mathbf A}{\mathbf x}_{{\kappa}}$. The measurements $\mathbf r$ is a componentwise probabilistic mapping of $\mathbf z$, i.e., $p({\mathbf r}|{\mathbf z})=\prod \limits_i p(r_i|z_i)$. To apply the GAMP algorithm, the i.i.d. Bernoulli Gaussian prior distribution is imposed for the vector ${\mathbf x}_{\kappa}$, i.e., $p({\mathbf x}_{\kappa})=\prod\limits_{i=1}^{I_{pd}}p({x}_{\kappa_i})$ and
\begin{align}\label{Bernoulli_Gaussian}
p({x}_{\kappa_i})=(1-\rho_{\kappa})\delta({x}_{\kappa_i})+\rho_{\kappa}{\mathcal {CN}}({x}_{\kappa_i};\mu_{\kappa},\sigma_{\kappa}^2),
\end{align}
where the sparsity rate $\rho_{\kappa}$, prior nonzero mean $\mu_{\kappa}$ and variance $\sigma_{\kappa}^2$ are all unknown. Let ${\boldsymbol \theta}=[\rho_{\kappa},\mu_{\kappa},\sigma_{\kappa}^2]^{\rm T}$. The expectation-maximization (EM) algorithm can be incorporated to iteratively learn ${\boldsymbol \theta}$ \cite{Vila}. For  details about the implementation, please refer to \cite{Jiang}. Since ${\mathbf A_{\kappa}}{\in}{\mathbb C}^{KLM{\times}I_{pd}}$, the computational complexity of the GAMP algorithm is $\mathcal{O}(KLN{I}_{pd})$. If  the GAMP algorithm is carried out directly, i.e., without dimension reduction, the computational complexity is $\mathcal{O}(KLNM_dM_{sp}M_r)$ and it is significantly higher than the proposed approach.

Let $\hat{\mathbf x}_{\kappa} = [\hat{x}_{\kappa}(0),..., \hat{x}_{\kappa}(i_{pd}), ...\hat{x}_{\kappa}(I_{pd}-1)]^{\rm T} \in {\mathbb C} ^{I_{pd} \times 1}$ denote the reconstructed vector. Based on $\hat{{\mathbf x}}_{\kappa}$, target detection can be then implemented to remove FAs in pre-detection vector $\tilde{{{\mathbf x}}}_{\kappa}$ and detect true targets. Let $\gamma_2$ be the detection threshold. Then, if $|\hat{x}_{\kappa}(i_{pd})|\geq \gamma_2$, a target is present. Otherwise, target is absent.

In order to evaluate the performance of the DR-GAMP algorithm, the reconstructed signal
\begin{align}\label{reconstsignal}
\hat{\mathbf s} = {\mathbf A}_{\kappa}\hat{{\mathbf x}}_{\kappa}
\end{align}
is used to compute the corresponding signal reconstruction error. As will be seen later, the strengths of harmonics in  $\hat{\mathbf s}$ are largely reduced.

It is worth noting that the  GAMP algorithm takes the quantization effects into consideration, and tries to find a  vector ${\mathbf x}_{{\kappa}}$  consistent with the observation model in (\ref{OnebitReceSigApprox}). As shown in \cite{meng1, zhu1MAP}, the  GAMP algorithm can be viewed as the iteration between the standard linear model (SLM) running AMP and the minimum mean squared error (MMSE) or MAP module. The MMSE or MAP module takes the quantization into account and iteratively refine the pseudo observations of the SLM. While for the SLM, the AMP takes the signal sparsity into account. As a result, the  GAMP algorithm has the powerful harmonic suppression capability.
\subsection{Considerations of the sampling frequency}

The sampling frequency is an important parameter. In this paper, we  apply the Nyquist sampling frequency, i.e., $f_s = B_r$, where $B_r$ is the bandwidth of the received signal before the one-bit quantization. From the linear processing point of view, the consideration of $f_s$ is as follows. Firstly, since the bandwidth of fundamental components  in the one-bit signal is still $B_r$ (refer to the spectrum analysis results in Section \ref{spectrum}), there is no frequency ambiguities for fundamental components at $f_s = B_r$. The pre-detection results can reveal grid points of true targets. Secondly, the spectrum aliasing\footnote{At $f_s = B_r$, it will cause frequency ambiguities for high-order harmonics. However, high-order harmonics are undesired components and can be removed in the second stage of the DR-GAMP approach.} caused by high-order harmonics has little impacts on target predetection.  The reason is that, main components are discrete and sparse in the frequency domain, as discussed in Section \ref{DRGAMP}.

From the respect of  nonlinear signal processing, recent works \cite{Stein1, Fettweis1, lmaz, Fettweis2} for wireless communications study the relationship between the bound of the Fisher information and sampling frequency. It is shown that the loss caused by the  one-bit ADC can be compensated via oversampling with respect to the Nyquist rate. Nevertheless, oversampling results in noise correlation and it is hard to design efficient algorithms. It is believed that one-bit radar benefits from oversampling. Characterizing the performance gain analytically is still an open problem for one-bit radar.
\section{Numerical Results}\label{NS}
In this section, substantial numerical experiments are provided from three aspects: Firstly,  effects of the one-bit quantization on target detection are evaluated. Secondly, the effectiveness of the proposed method on reconstructing targets and suppressing FAs caused by high-order harmonics are investigated. Thirdly, the target detection performance is presented and performance comparisons between one-bit radar and the conventional radar are provided.

The variance of the additive complex noise $2\sigma_w^2$ is assumed to be available since this information can be estimated through a training process in the built-in self-test (BIST) stage of radar. Here we set $2\sigma_w^2=1$. To perform the DR-GAMP approach, the $\eta$th statistics (denoted by $x_\eta$ in (\ref{gamma1set})) in the three domains are set as the $\lfloor 0.75R_d \rfloor$th, $\lfloor 0.75R_{sp} \rfloor$th and $\lfloor 0.75R_{r} \rfloor$th elements of the ordered lists of reference cells, where $R_d$, $R_{sp}$ and $R_r$ are the corresponding numbers of reference cells. For the scale factor $\alpha_{OS}$ in (\ref{gamma1set}), it is set according to the noise false alarm $P_{{\rm FA},w}$ and is given in the corresponding simulations.

\subsection{Effects of the one-bit quantization on target detection}
In this subsection, we firstly validate the assumption (\ref{Gau_ass}) via numerical experiments. Then, attenuations of 3-order harmonics with respect to the fundamental frequencies, FAs caused by 3-order harmonics and SNR losses caused by the one-bit quantization are discussed. In the frequency domain, define the SNR of a fundamental component or a high-order harmonic (\ref{Gau_ass}) as
\begin{align}\label{harmonidef}
{\rm SNR}_q=|A_q|^2/(2\sigma_{w_f}^2).
\end{align}
\subsubsection{Validation of the noise  statistics in (\ref {Gau_ass})}
\begin{figure}[h!t]
\centering
\includegraphics[width=3.5in]{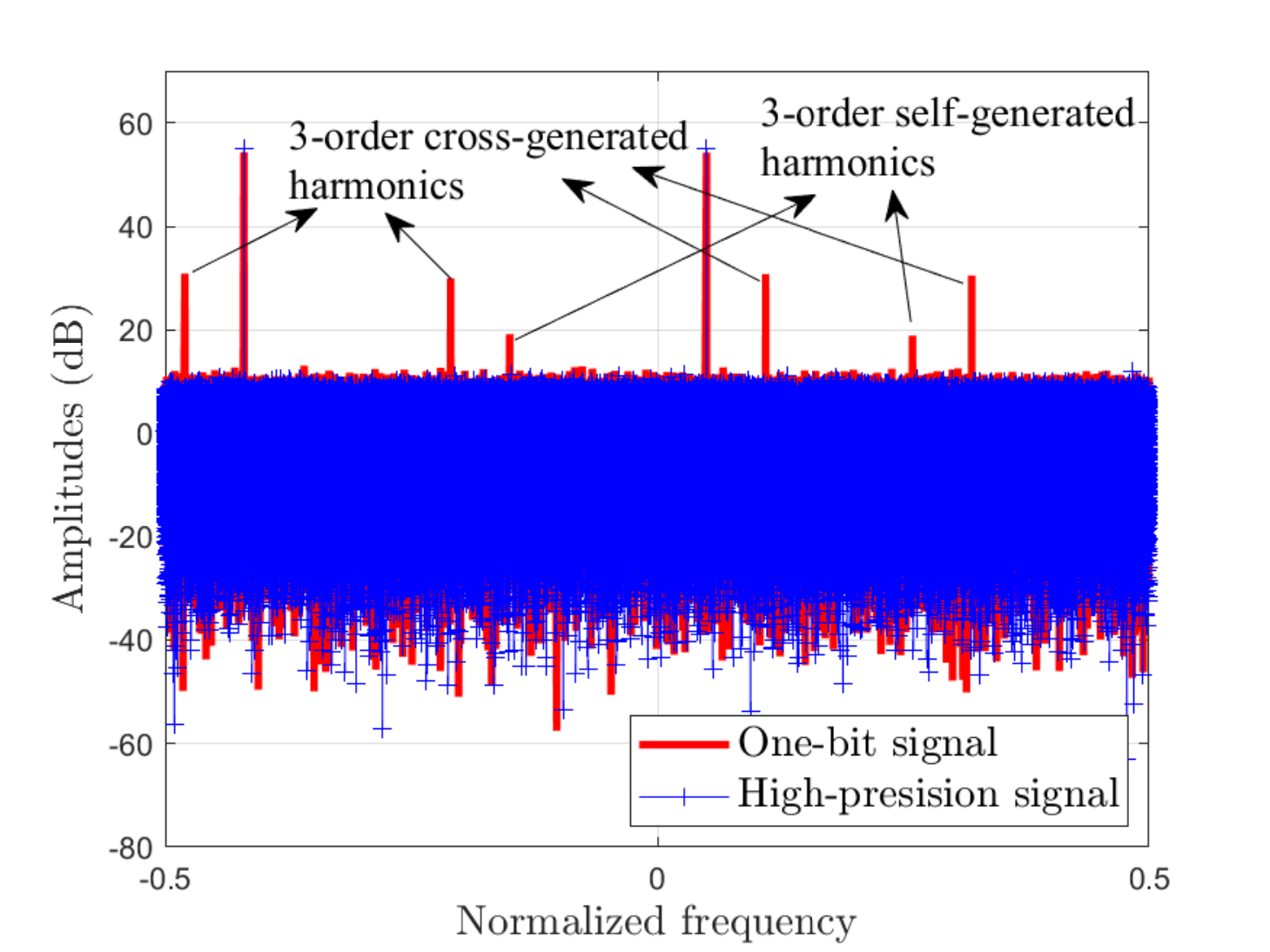}
\caption{Spectrums of the high-precision and one-bit quantization signals. Before the one-bit quantization, the received SNRs of the two targets are ${\rm SNR}_1={\rm SNR}_2 = -5$ dB. Both the number of observations and the number of FFT points are $10^6$. Note that because $f_s = B_r$, frequency ambiguities exist for several 3-order harmonics. }
\label{Spectrum_5dB}
\end{figure}

\begin{figure}[h!t]
\centering
\includegraphics[width=6in]{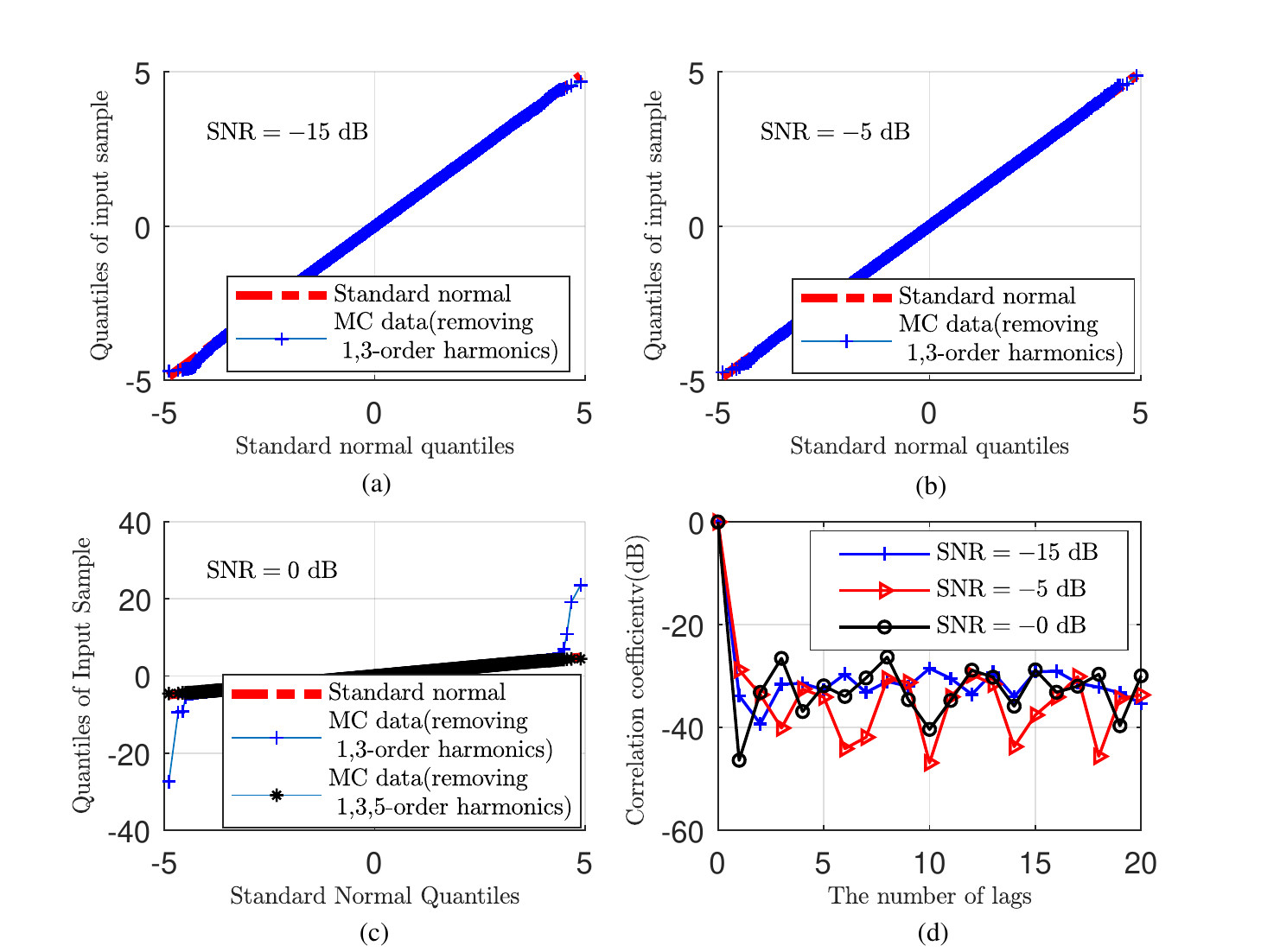}
\caption{QQplots and autocorrelation coefficient curves for checking the Gaussianity of data after FFT. (a) ${\rm SNR} = -15$ dB. (b) ${\rm SNR} = -5$ dB. (c) ${\rm SNR} = 0$ dB. (d)  Autocorrelation coefficients of noise cells for the three SNR cases.}
\label{realPartQQ}
\end{figure}
In (\ref {Gau_ass}), we assume that $w(f)$ is the AWGN. Without loss of generality, we take the fast time domain for example and consider the received one-bit signal $v_{qC}(t)$ in (\ref {fastSig}). Assume that the number of targets is $P=2$. The sampling frequency is $100$ MHz and the normalized frequencies of the two targets are $f_{r,1}=0.4$ and $f_{r,2}=0.05$, respectively.  The observation time duration is $10$ ms and the corresponding number of samples is $N = 10^6$.
The spectrum of the one-bit signal is shown in Fig. \ref{Spectrum_5dB}. It is shown that energies for both fundamental frequencies and high-order harmonics are focused in the frequency domain at certain frequency cells.

In order to analyze the noise statistics, the real part of ${{\mathcal F}(v_{qC}(t))}$, without loss of generality, is considered. Removing frequency cells of the fundamental frequencies and 3-order hamrmonics, QQplots\footnote{A QQplot is a visual inspection tool for checking the Gaussianity of the data. In a QQplot, deviation from a straight line is an evidence of non-Gaussianity.} and autocorrelation coefficient curves are provided in Fig. \ref{realPartQQ} (a)-(d) for different SNR cases. For the cases of ${\rm SNR} = -15$ dB and ${\rm SNR} = -5$ dB shown in Fig. \ref{realPartQQ} (a) and (b), the QQplots of the remaining frequency cells are approximately a straight line. This means that the noise is approximately Gaussian. Nevertheless, for the case ${\rm SNR} = 0$ dB shown in Fig. \ref{realPartQQ} (c), several large values deviate from the straight line. The reason is that at ${\rm SNR} = 0$ dB, the 5-order harmonics can not be ignored again. If we further remove cells of 5-order harmonics, the QQplot is back to a straight line in Fig. \ref{realPartQQ} (c) and hence, the noise is approximately Gaussian as well. Autocorrelation coefficients of noise frequency cells are shown in Fig. \ref{realPartQQ} (d). Clearly, the noise in the frequency domain is almost independent. Hence, $w(f)$ in (\ref{Gau_ass}) can be modeled as the AWGN.

\subsubsection{FAs caused by high-order harmonics}
\begin{figure}[h!t]
\centering
\includegraphics[width=3.5in]{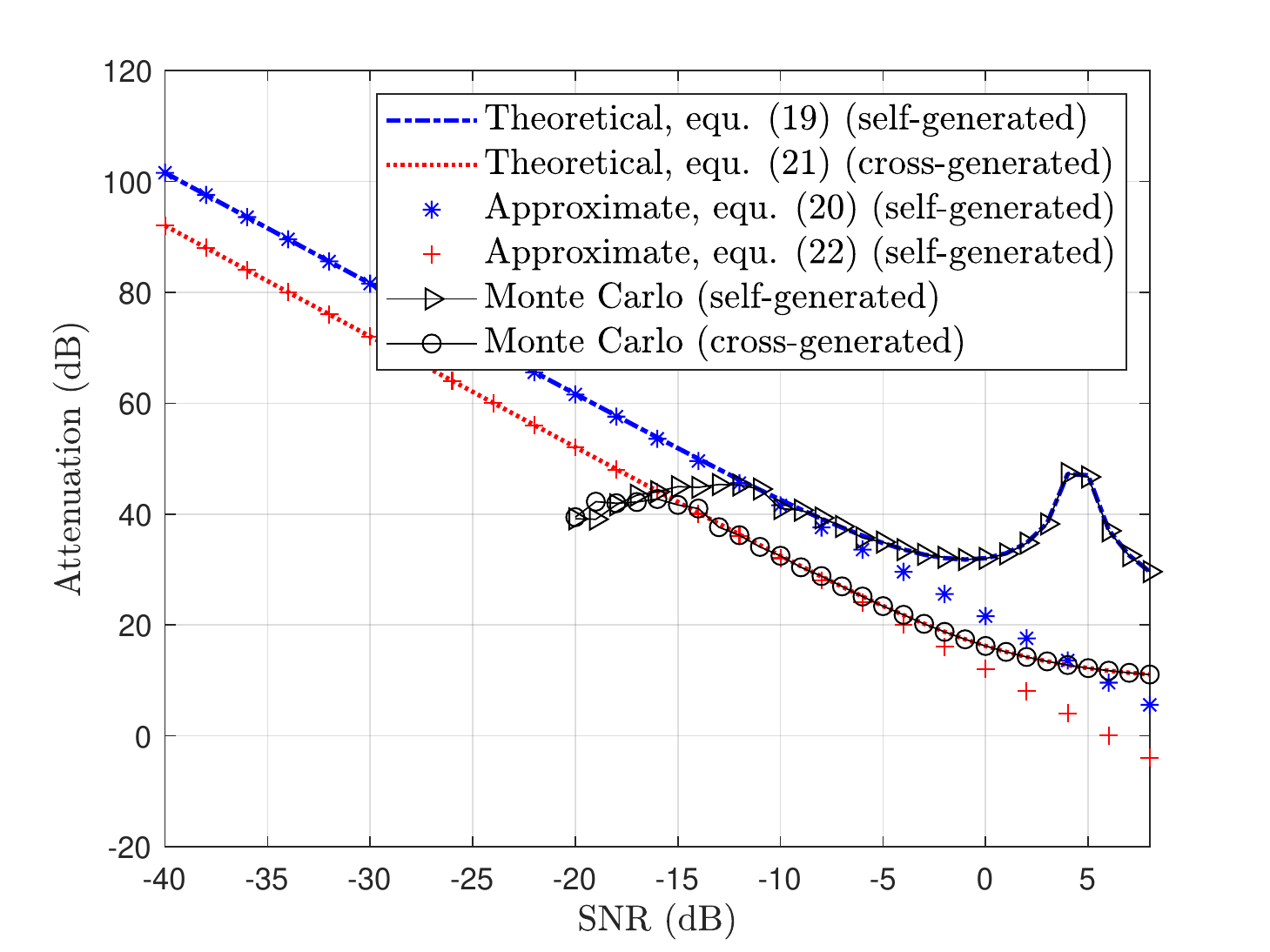}
\caption{Attenuations of 3-order harmonics with respect to the fundamental component.}
\label{3_orderAttenuation_pdf}
\end{figure}

We still consider the  scenario with two targets and set ${\rm SNR}_1={\rm SNR}_2$. The frequencies of the two targets are denotes as $f_{r,1}$ and $f_{r,2}$, respectively. The attenuations of the self-generated and cross-generated 3-order harmonics with respect to the fundamental component are shown in Fig. {\ref{3_orderAttenuation_pdf}}. It can be seen that when ${\rm SNR}<-10$ dB, the theoretical results (\ref{m-order-res_furth1}) and (\ref{cross-order-res_furth1}) are consistent with their approximations (\ref{m-order-res_furth2}) and (\ref{cross_res}). In addition, when SNR satisfies  $-13~{\rm }$  dB$\leq$ ${\rm SNR}\leq 8~{\rm }$ dB, the Monte Carlo (MC) results are consistent with (\ref{m-order-res_furth1}) and (\ref{cross-order-res_furth1}) as well. This simulation validates the theoretical analysis in Section \ref{spectrum}.

In addition, Fig. {\ref{3_orderAttenuation_pdf}} shows that when the SNR is low, the attenuations are very large and both the self-generated and cross-generated 3-order harmonics can be ignored. As SNR increases, the attenuations becomes smaller and the strength of the 3-order harmonics can not be ignored again. For example, when ${\rm SNR}=-5$ dB, the SNRs of the 3-order cross-generated and self-generated harmonics are about $23.6$ dB and $34.8$ dB lower than that of the fundamental component, respectively. The 3-order harmonics will result in FAs in many application scenarios. Consider a set of typical parameters (details will be shown later in Table \ref{tab:Radar_parameters}): $K=200$, $L=24$ and $N=1000$. The ideal digital integration gain is $10{\rm log}(KLN)=66.8$ dB (without considering the processing loss). Then, after integration, the SNRs of the 3-order cross-generated and self-generated harmonics are about $38.2$ dB and $27.0$ dB, respectively. As a consequence, the 3-order harmonics can not be ignored. Since 3-order harmonics are target-like (as shown in Fig. \ref{Spectrum_5dB}), they will result in FAs.

Furthermore, since the corresponding frequency range of 3-order harmonics is partly overlap with that of true targets (no matter how large the sampling frequency $f_s$ is selected), linear processing methods are difficult to completely remove FAs caused by harmonics. For example, there are two 3-order harmonics with their frequencies\footnote{The frequency locations of the 3-order harmonics are $-3f_1$, $-3f_2$, $-2f_1-f_2$, $-2f_2-f_1$, $2f_1-f_2$ and $2f_2-f_1$, respectively. In \cite{Pascazio1998}, oversampling are used to separate harmonics from the fundamental frequencies. However, the last two harmonics may lie in the range of  targets and can not be removed via linear processing approaches.}  of $2f_{r,1}-f_{r,2}$ and $2f_{r,2}-f_{r,1}$, and the two values may be equal to that of true targets. Hence, the two harmonics can not be removed, otherwise true targets may fail to detect.

\subsubsection{The SNR loss of the one-bit quatization}
For the one-bit quantization, fundamental frequencies suffer SNR losses comparing with the high-precision quantization especially for scenarios with strong targets. For a given target, the SNR loss is defined as ${\rm SNR}_c-{\rm SNR}_{\rm ob}$ (expressed in dB), where ${\rm SNR}_c$ and ${\rm SNR}_{\rm ob}$ denote the target SNRs (refer to (\ref{harmonidef})) in the conventional system and one-bit radar.
\begin{figure}[h!t]
\centering
\includegraphics[width=3.5in]{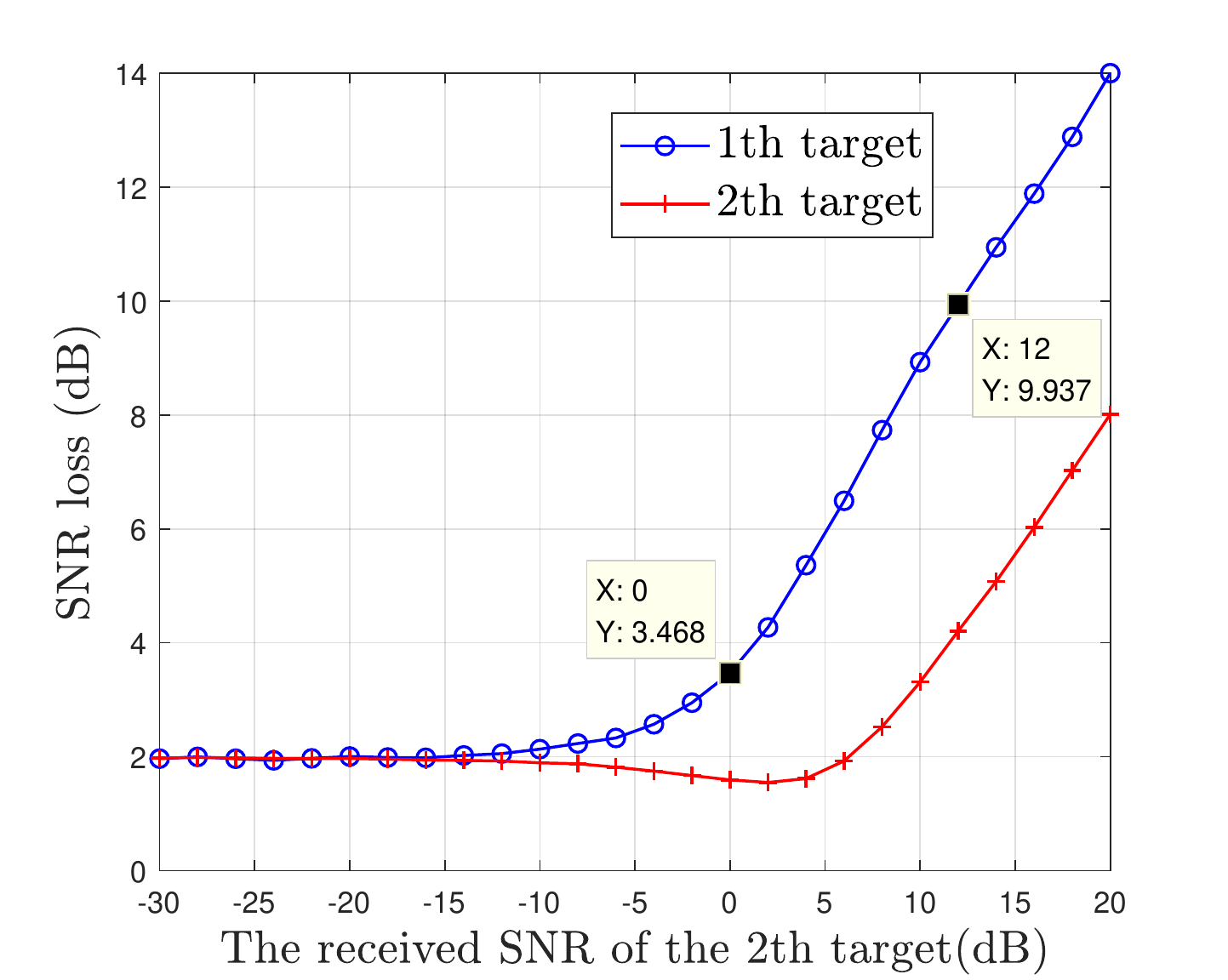}
\caption{SNR losses caused by the one-bit quantization in a scenario with two targets. The blue solid line with the circle marker and the red solid line with the plus-sign marker denote the SNR losses for the 1th and 2th target, respectively.}
\label{SNRloss}
\end{figure}

We again consider the scenario with two targets. For the first target, we fix its SNR as ${\rm SNR}_1 = -30$ dB. For the second target, its SNR  varies from $-30$ dB to $20$ dB. The SNR losses caused by the one-bit quantization are calculated and results are shown in Fig. \ref{SNRloss}. When the SNR of the second target satisfies  ${\rm SNR}_2<-10$ dB, the SNR losses are both about $2$ dB. As ${\rm SNR}_2$ increases from $-10$ dB to $20$ dB, the SNR loss of the first target increases from $2$ dB to $14$ dB. At ${\rm SNR}_2=0$ dB, the SNR loss of the first target is about $3.5$ dB which is generally acceptable in practice. However, when ${\rm SNR}_2>12$ dB, the SNR loss of the first target is about $10$ dB and this loss will significantly degrade the detection performance of targets with low SNRs.

The SNR losses indicate that due to the application of the one-bit quantization, the dynamic range of one-bit radar is reduced. A possible solution to this problem is to adopt the time varying threshold \cite{Wangpu, Li2016, HL2018}.
\subsection{Effectiveness of the DR-GAMP approach}
In this subsection, effectiveness of the DR-GAMP approach is investigated under both on-grid and off-grid cases. Firstly, the benefit of dimension reduction in the DR-GAMP approach is elucidated. Then, the excellent performance of the DR-GAMP approach in terms of high-order harmonic suppression is demonstrated. To evaluate the reconstruction performance, the normalized mean squared error (NMSE)
\begin{align}
{\rm NMSE}=20\log \frac{\|\hat{\mathbf s}-{\mathbf A}{\mathbf x}\|_2}{\|{\mathbf A}{\mathbf x}\|_2}
\end{align}
is used, where $\hat{\mathbf s}$ is defined in (\ref{reconstsignal}).
\subsubsection{Benefit of dimension reduction}
\begin{figure}[htb]\centering \subfigure[]{\begin{minipage}[t]{0.47\linewidth}\centering\includegraphics[width=3.3in]{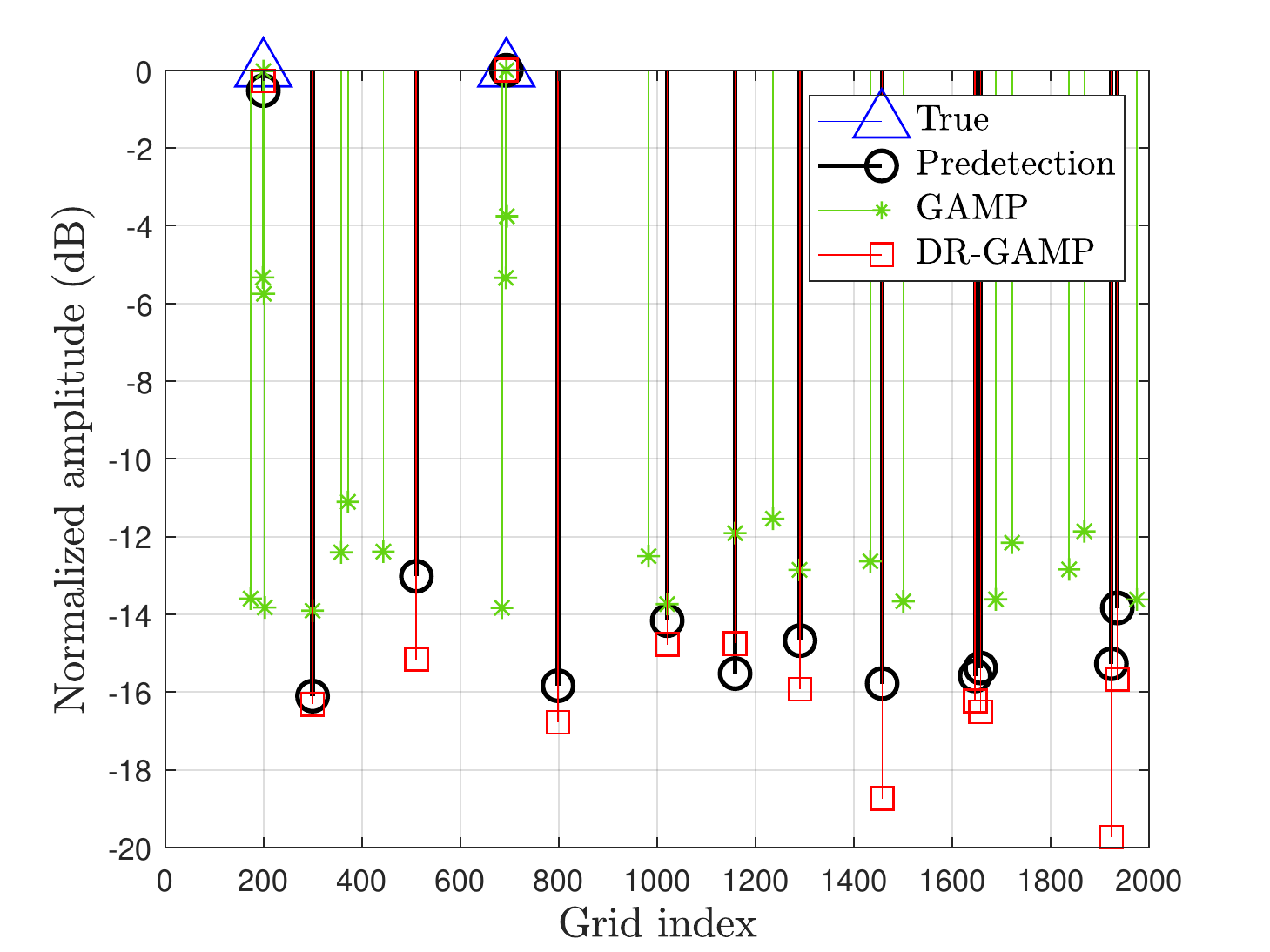}
\label{OngGridRecRes}
\end{minipage}}
\subfigure[]{\begin{minipage}[t]{0.45\linewidth}\centering\includegraphics[width=3.3in]{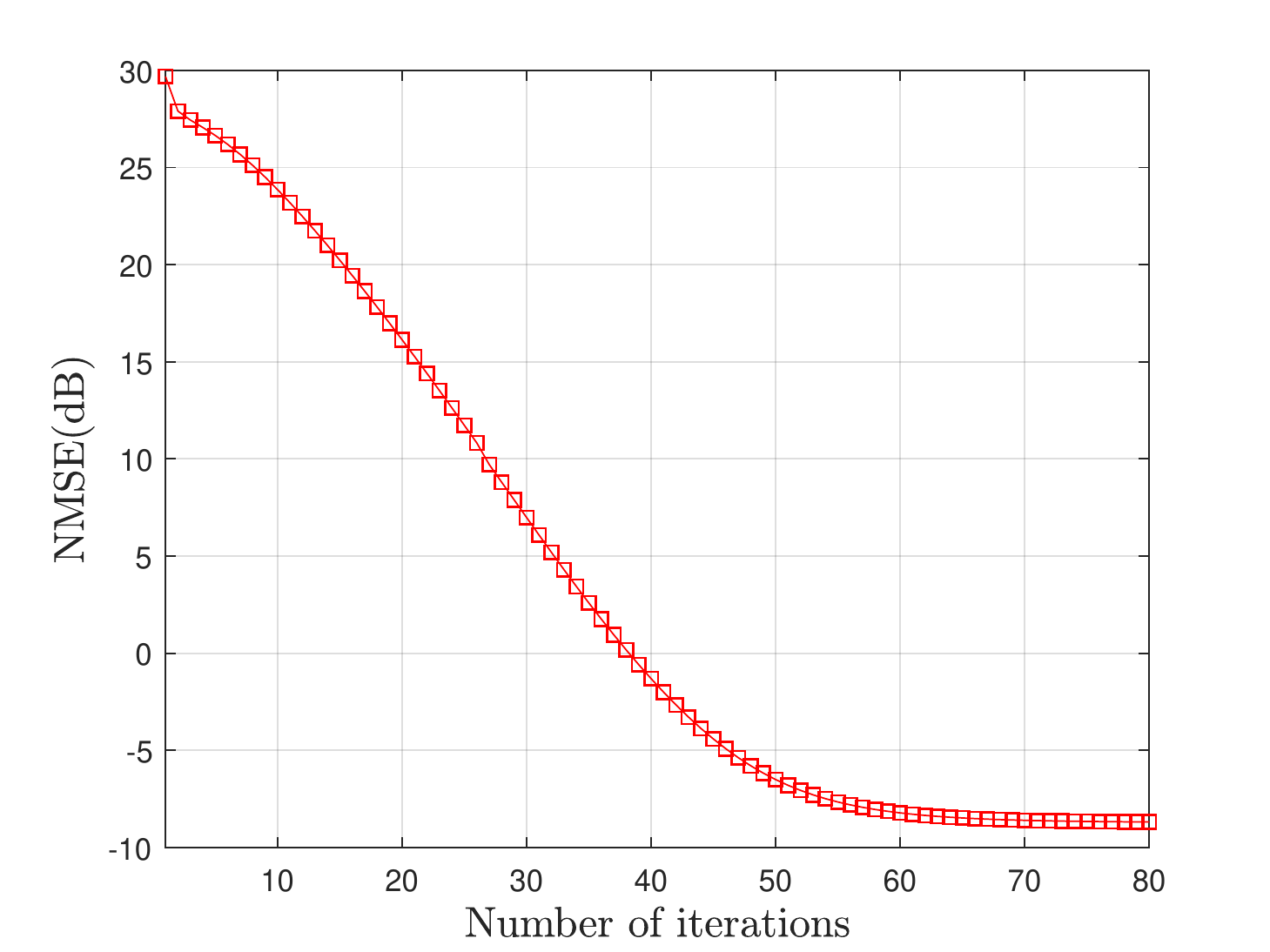}
\label{Ongrid_MSE}
\end{minipage}}
\centering\caption{The on-grid case. Fig. \ref{dimensionRedOngridcase} (a) presents the normalized amplitudes of $\mathbf x$, $\tilde{\mathbf x}_{\kappa}$, $\hat{{\mathbf x}}_{\rm GAMP}$ and $\hat{{\mathbf x}}_{\kappa}$, where $\hat{{\mathbf x}}_{\rm GAMP}$ denotes the recovery vector obtained by directly performing the GAMP algorithm without dimension reduction.  The normalized amplitudes are calculated such that the maximum amplitude for each of the four vectors is normalized to $0$ dB. For $\hat{{\mathbf x}}_{\rm GAMP}$, only the normalized amplitudes above $-15$ dB are plotted for clarity. In the simulation, the maximal amplitudes of $\mathbf x$, $\tilde{\mathbf x}_{\kappa}$, $\hat{{\mathbf x}}_{\rm GAMP}$ and $\hat{{\mathbf x}}_{\kappa}$ are $-5.0$ dB, $23.4$ dB, $38.3$ dB and $-5.0$ dB, respectively.  Fig. \ref{dimensionRedOngridcase} (b) presents the NMSE of the DR-GAMP approach. }\label{dimensionRedOngridcase}
\end{figure}

For convenience, we still take the fast time domain for example and consider a scenario with two targets. The SNRs of the two targets satisfy  ${\rm SNR}_1={\rm SNR}_2=-5$ dB. In order to investigate the recovery performance with some grid points corresponding to the noise FAs, we set a relatively small scale factor, and here $\alpha_{OS}=8.9$ dB. At $\alpha_{OS}=8.9$ dB, the noise FA rate is about 0.006.  In the simulation, the number of samples is $N=1000$. The bandwidth of the received signal and the sampling frequency are both 100 MHz, i.e., $B_r=f_s=100$ MHz\footnote{We choose $N=1000$ for two reasons: (1) The GAMP algorithm can be directly implemented since the dimension is not huge. (2) The harmonics can be neglected and we thus focus on the benefit of dimension reduction. As shown in Fig. \ref{3_orderAttenuation_pdf}, the attenuation of the 3-order cross-generated harmonics compared to the fundamental frequency is at least $23$ dB. For one-bit quantization, Fig. \ref{SNRloss} shows the SNR loss (compared to the conventional radar) is at least $2$ dB. The integration gain is $10\log N=30$ dB. Thus, the SNRs of the 3-order cross-generated harmonics are about $-5-2+30-23=0$ dB.}. The overgriding factor in (\ref{defovergriding}) is set as $r_a=2$, thus $M_r=r_aN=2000$ and the grid interval is $f_s/M_r=0.05$ MHz.

\begin{figure}[h!t]
\centering
\includegraphics[width=6.0in]{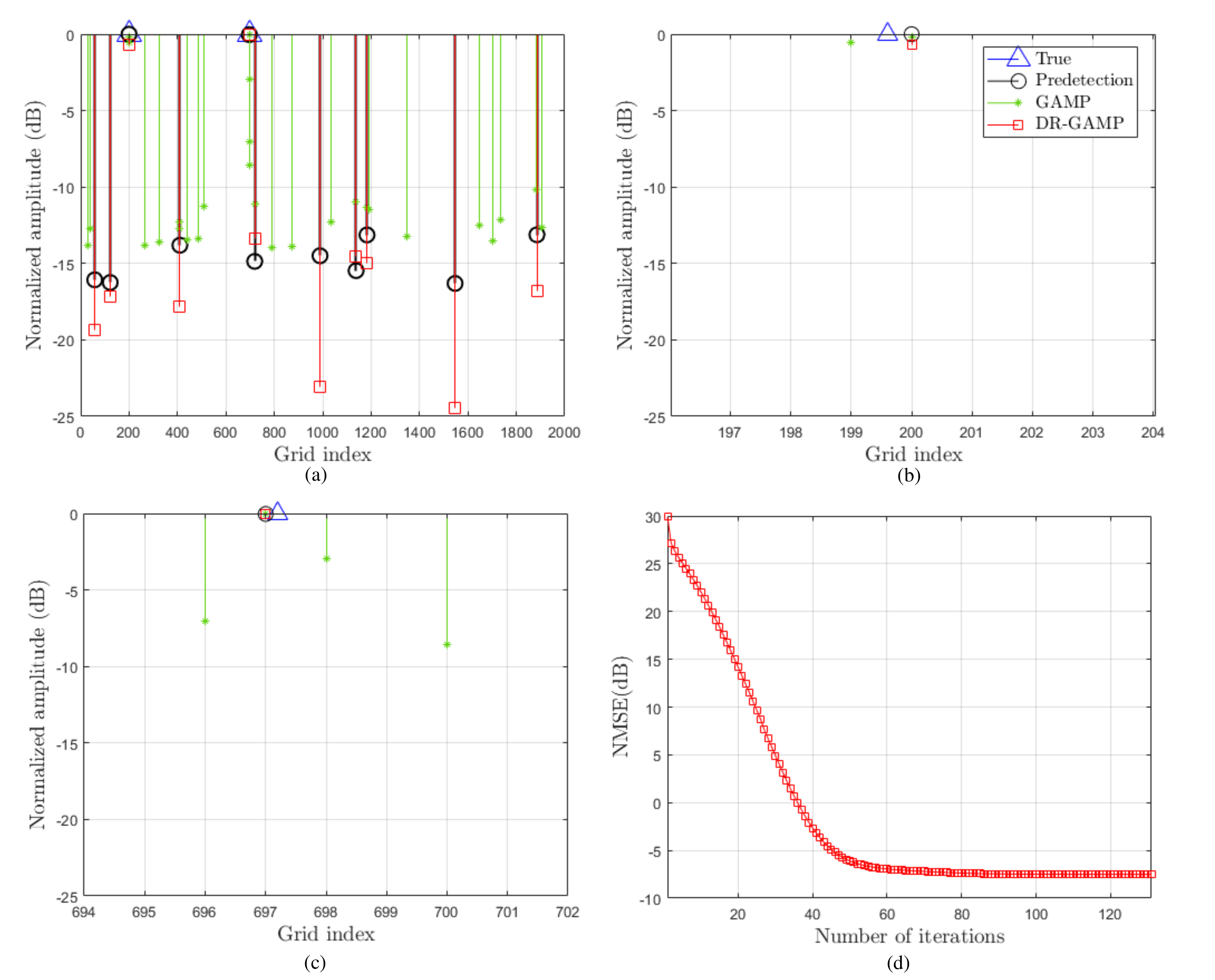}
\caption {The off-grid case. Fig. \ref{Offfgrid_combine} (a) presents the normalized amplitudes of $\mathbf x$, $\tilde{\mathbf x}_{\kappa}$ $\hat{{\mathbf x}}_{\rm GAMP}$, and $\hat{{\mathbf x}}_{\kappa}$. In the simulation, the maximal amplitudes of $\mathbf x$, $\tilde{\mathbf x}_{\kappa}$, $\hat{{\mathbf x}}_{\rm GAMP}$ and $\hat{{\mathbf x}}_{\kappa}$ are $-5.0$ dB, $23.7$ dB, $38.4$ dB and $-5.4$ dB, respectively. Fig. \ref{Offfgrid_combine} (b) and Fig. \ref{Offfgrid_combine} (c) present the fine details around the first and second target, respectively. Fig. (d) plots the NMSE of the DR-GAMP approach.}
\label{Offfgrid_combine}
\end{figure}

\begin{itemize}
  \item The on-grid scenario: The frequencies of the two targets are $f_{r,1}= -40.1$ MHz and $f_{r,2}= -15.4$ MHz, respectively. Results are shown in Fig. \ref{dimensionRedOngridcase}. Fig. \ref{dimensionRedOngridcase} (a) shows that several false components around true targets (their amplitudes are only $4\sim 6$ dB lower than the true targets) are reconstructed by the GAMP algorithm, which will lead to FAs. In contrast, the DR-GAMP approach recovers the true targets without false components, and its NMSE is about $-8$ dB, as shown  in Fig. \ref{dimensionRedOngridcase} (b). The above results demonstrate the benefit of dimension reduction.
  \item The off-grid scenario: The frequencies of the two targets are $f_{r,1}= -40.07$ MHz and $f_{r,2}= -15.19$ MHz. The biases of the two frequencies from their nearest grid points (the corresponding frequencies are $-40.05$ MHz and $-15.20$ MHz, respectively) are $0.02$ MHz and $0.01$ MHz, respectively. Results are shown in Fig. \ref{Offfgrid_combine}. The phenomenon is basically the same as that in the on-grid scenario. The DR-GAMP approach still works well and reconstructs the targets on the grid points which are nearest to the true frequencies.
\end{itemize}

\subsubsection{High-order harmonic suppression}

We consider the on-grid and off-grid scenarios with two targets. Simulation parameters are shown in Table \ref{tab:Radar_parameters}. Here, we mainly focus on high-order harmonic FAs and set a relative high pre-detection threshold $\gamma_1$. The scale factor $\alpha_{OS}$ in $\gamma_1$ is set as 10.6 dB. The SNR of the two targets before the one-bit quantization are the same and $\rm {SNR_1=SNR_2=-7}$ dB.

\newcommand{\tabincell}[2]{\begin{tabular}{@{}#1@{}}#2\end{tabular}}
\begin{table}[!t]
  \centering
  \scriptsize
  \caption{Simulation parameters}
  \label{tab:Radar_parameters}
  \begin{tabular}{ll}
    \\[-2mm]
    \hline
    \hline\\[-2mm]
    { \small Parameters}&\qquad {\small Value}\\
    \hline
    \vspace{1mm}\\[-3mm]
    Carrier frequency $f_c$   &   24GHz\\
    \vspace{1mm}
    Frequency modulation slope $\mu$          &  $10^{13}$Hz/s\\
    \vspace{1mm}
     Pulse repeat interval $T_I$         &  $2{\times }10^{-5}$s\\
    \vspace{1mm}
    Bandwidth $B_r$          &  100MHz\\
    \vspace{1mm}
    Sampling frequency $f_s$         &  100MHz\\
    \vspace{1mm}
    Number of pulses $K$          &  200\\
    \vspace{1mm}
    Number of antenna elements $L$          &  24\\
    \vspace{1mm}
    Antenna element spacing $d$  &  0.00625m\\
    \vspace{1mm}
    Number of fast time samples $N$          &  1000\\
    \vspace{1mm}
    Spatial window          &  Taylor\\
    \vspace{1mm}
    Slow time window and peak sidelobe           &  Chebyshev, $-60$ dB\\
    \vspace{1mm}
    Fast time window and peak sidelobe           &  Chebyshev, $-60$ dB\\
    \hline
    \hline
  \end{tabular}
\end{table}

For the on-grid  scenario, the two targets share both the same zero spatial frequencies, i.e., $f_{sp,1} = f_{sp,2}=0$. Beat frequencies and Doppler shifts of the two targets are $f_{r,1}=-40$ MHz, $f_{r,2}= -3$ MHz, $f_{d，1} = 2$ KHz and $f_{d，2} = 7$ KHz, respectively. Note that the above parameter settings make the high-order harmonic FAs appear merely in the $c_{sp}$th spatial cell with its spatial frequency of $f_{sp,c_{sp}}=0$. The overgriding factor $r_a$ in (\ref{defovergriding}) is set as $r_a=2$.

\begin{figure}[h!t]
\centering
\includegraphics[width=6.0in]{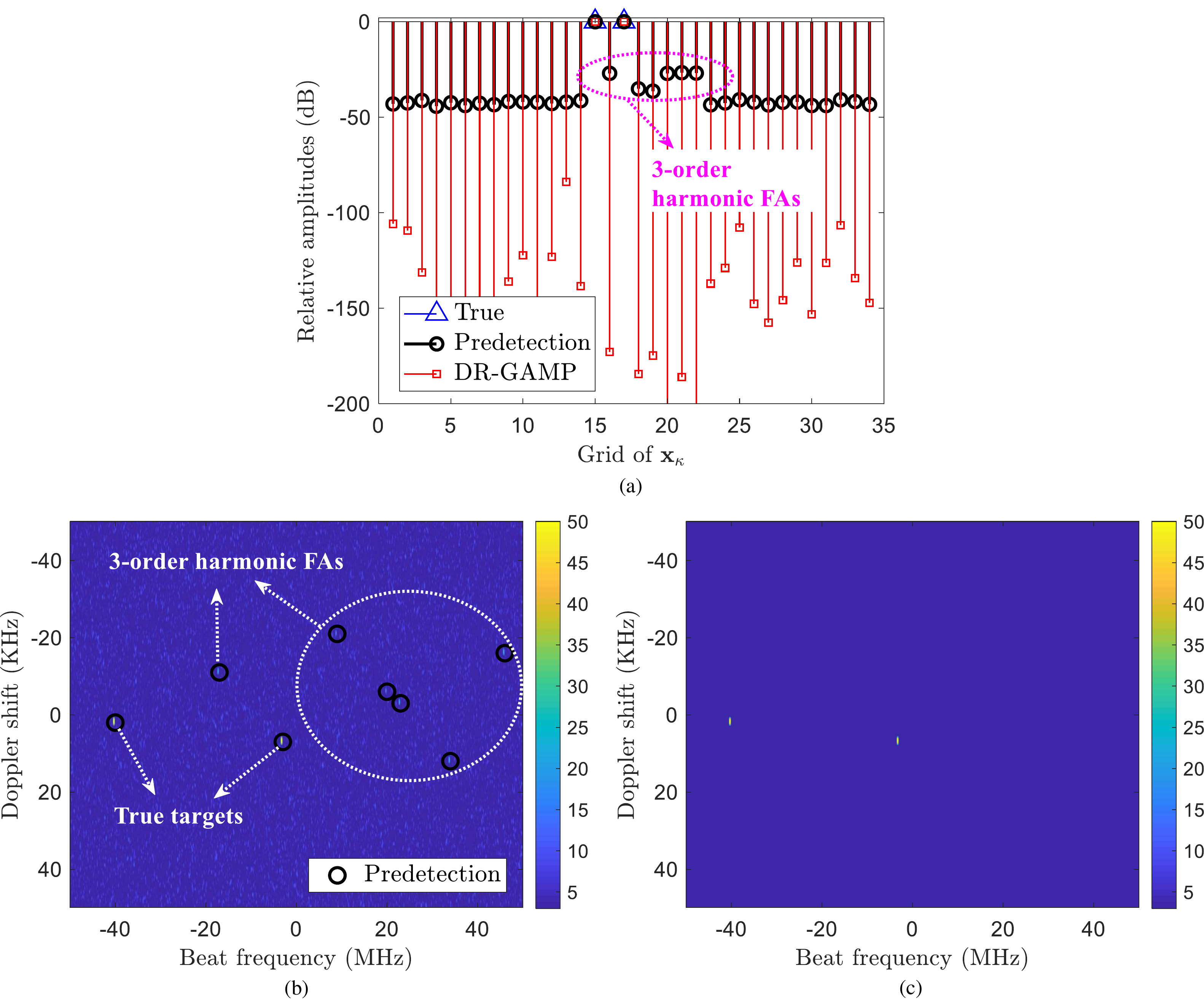}
\caption {The on-grid  scenario. (a) Normalized amplitudes of ${\mathbf x}_{\kappa}$, $\tilde{{{\mathbf x}}}_{\kappa}$ and $\hat{{\mathbf x}}_{\kappa}$. The maximal amplitudes of ${\mathbf x}_{\kappa}$, $\tilde{{{\mathbf x}}}_{\kappa}$ and $\hat{{\mathbf x}}_{\kappa}$ are $-7.0$ dB, $55.4$ dB and $-7.0$ dB, respectively. (c) Pre-detection results are plotted on the range-Doppler map at the spatial cell with $f_{sp}=0$. (d) Reconstructed range-Doppler map (based on the reconstructed signal in (\ref{reconstsignal})) at the spatial cell with its spatial frequency of 0.}
\label{firstsce}
\end{figure}

\begin{figure}[h!t]
\centering
\includegraphics[width=6.0in]{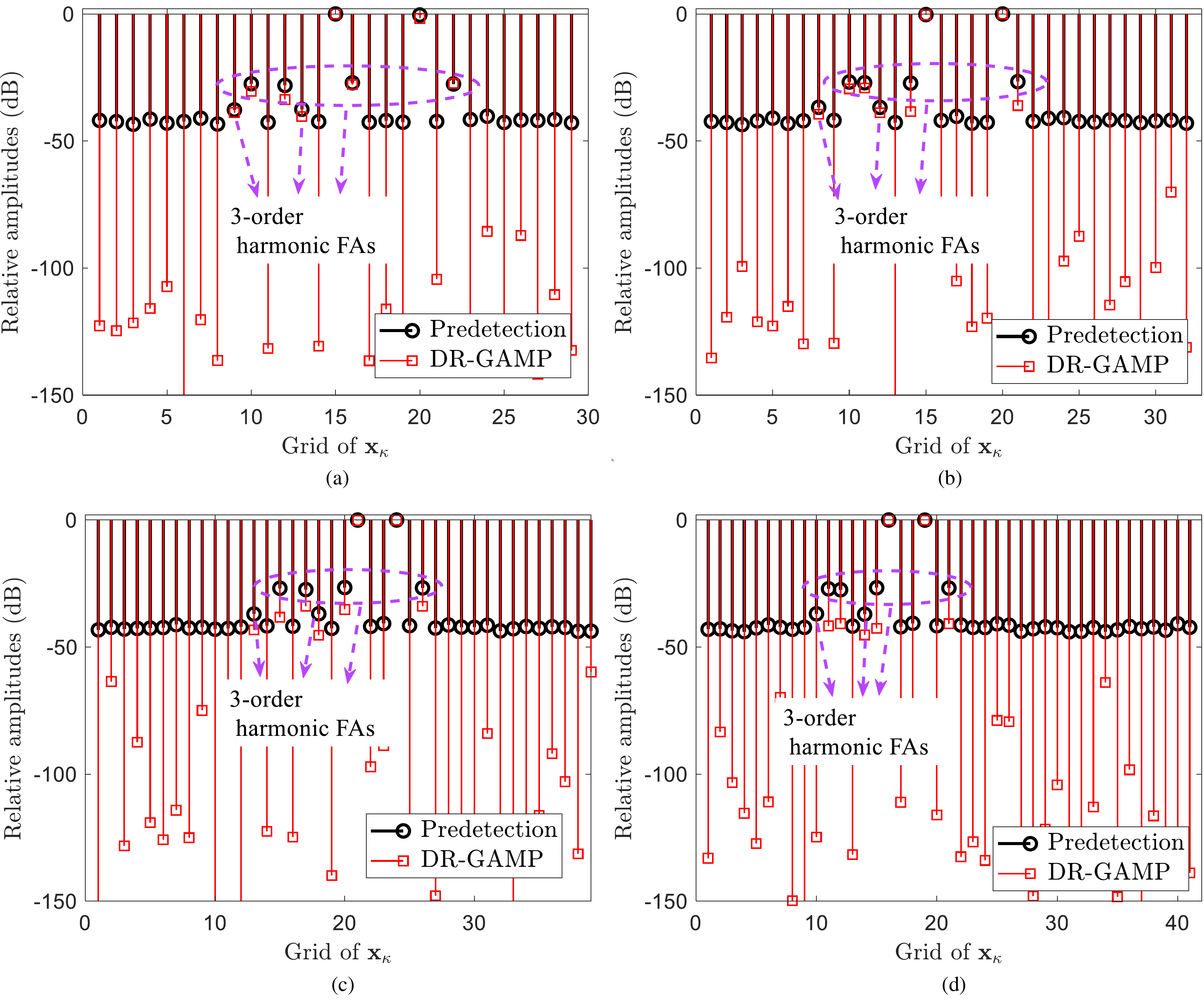}
\caption {The off-grid  scenario. Fig. \ref{OffgridCase} (a), \ref{OffgridCase} (b), \ref{OffgridCase} (c) and \ref{OffgridCase} (d) present the normalized amplitudes of $\tilde{{{\mathbf x}}}_{\kappa}$ and $\hat{{\mathbf x}}_{\kappa}$ at $r_a=1,2,3,4$, respectively. The maximal amplitudes of $\tilde{{{\mathbf x}}}_{\kappa}$ and $\hat{{\mathbf x}}_{\kappa}$ are $54.2$ dB and $-9.9$ dB ($r_a=1$), $54.9$ dB and $-8.4$ dB ($r_a=2$), $55.1$ dB and $-7.6$ dB ($r_a=3$), $55.2$ dB and $-7.3$ dB ($r_a=4$), respectively.}
\label{OffgridCase}
\end{figure}

For the off-grid  scenario, targets parameters are set as follows: $f_{d,1} = 2.1$ KHz, $f_{d,2} = 7.35$ KHz, $f_{sp,1} = 0.0125$ Hz, $f_{sp,2} = 0.0354$ Hz, $f_{r,1} = -40.015$ MHz, $f_{r,2} = -3.06$ MHz. We investigate the performance under $r_a=1,2,3,4$, respectively. For each $r_a$, the biases between the true frequencies of targets in the three domains and their nearest grid points are given in Table \ref{tab:Offgrid_parameters}. For the two scenarios, numerical results are shown in Fig. \ref{firstsce} and Fig. \ref{OffgridCase}, respectively.

\begin{table}[!t]
\begin{threeparttable}
  \centering
  \scriptsize
  \caption{Parameter settings for the off-gird scenario.}
  \label{tab:Offgrid_parameters}
  \begin{tabular}{|c|c|c|c|c|c|c|c|c|c|}
    \\[-7mm]
    \hline\\[-4mm]
    $r_a$   &  $\!\!\! {\rm PRF}/M_d $ (KHz) \!\!\!\!\!&  $\!\!\!{1}/M_{sp}$ (Hz) &  $\!\!\!{f_s}/M_{r}$ (MHz)\!\!\!\! & $\!\!\!\Delta f_{d,1}$\!\!\! (KHz)\!\!\! & $\!\!\!\Delta f_{d,2}$ (KHz)\!\!\! & $\!\!\!\Delta f_{sp,1}$ (Hz)\!\!\!& $\!\!\! \Delta f_{sp,2}$ (Hz)\!\!\!\! & $\!\!\! \Delta f_{r,1}$ (MHz)\!\!& $\!\! \Delta f_{r,2}$ (MHz)\!\!\\
    \hline
    $1$   &  $0.5$ &  $0.0417$ &  $0.1$ & $0.1$  & $0.15$  & $0.0125$  & $0.0063$  & $0.015$  & $0.04$\\
    $2$   &  $0.25$ &  $0.0208$ &  $0.05$ & $0.1$  & $0.1$  & $0.0083$  & $0.0063$  & $0.015$  & $0.01$\\
    $3$   &  $0.167$ &  $0.139$ &  $0.033$ & $0.067$  & $0.0167$  & $0.0014$  & $0.0063$  & $0.015$  & $0.0067$\\
    $4$   &  $0.125$ &  $0.0104$ &  $0.025$ & $0.025$  & $0.025$  & $0.0021$  & $0.0042$  & $0.01$  & $0.01$\\
    \hline
  \end{tabular}
  \begin{tablenotes}
\item[1] $\Delta f$ denotes the bias between the true frequency value of a target and its nearest grid point.
\end{tablenotes}
\end{threeparttable}
\end{table}

\begin{itemize}
  \item The on-grid scenario:  In Fig. \ref{firstsce} (a), normalized amplitudes of ${\mathbf x}_{\kappa}$, $\tilde{{{\mathbf x}}}_{\kappa}$ and $\hat{{\mathbf x}}_{\kappa}$ are presented. It can be seen that after predetection, $34$ PTs are detected, including $2$ true targets, $6$ $3$-order harmonic FAs and $26$ noise FAs. Then, applying the DR-GAMP algorithm, all of the $3$-order harmonics are suppressed significantly.  Range-Doppler maps are provided in Fig. \ref{firstsce} (b) and \ref{firstsce} (c) for the one-bit signal and reconstructed signal at the spatial cell with its spatial frequency of 0. Results show that 3-order harmonics in the one-bit signal are strong, while their strengths in the DR-GAMP reconstructed signal are negligible. That is, the proposed approach suppresses the 3-order harmonics significantly.
  \item The off-grid scenario: Simulation results are plotted in Fig. \ref{OffgridCase}. The ability of high-order harmonic suppression improves as $r_a$ increases for the DR-GAMP approach. For $r_a=1$, the DR-GAMP approach is almost ineffective to suppress harmonics. As $r_a$ increases to 4, the harmonic suppression ratio improves effectively.
\end{itemize}

\subsection{Performance comparisons between one-bit and conventional radar}
In this subsection, the detection performance of one-bit radar is compared with that of the conventional radar under the off-grid case. For the conventional radar, the received signal is digitized using the conventional high-precision ADCs, where the quantization error can be omitted. The 3-D FFT is applied to perform the coherent integration and then, the OS CFAR is adopted directly to implement target detection.

Two scenarios are considered:
\begin{itemize}
  \item Scenario 1: $10$ targets with identical SNRs are added for each simulation trial. The SNRs of targets vary from $-38$ dB to $-28$ dB ($\rm {SNR} = -38, -37, ..., -28$).  For each value of SNR, the number of MC trials is 500.
  \item Scenario 2:  A new strong target with its SNR of $0$ dB is added in the scenario 1.
\end{itemize}

For each trial, all targets are randomly added in the spatial, slow time, and fast time domain. The number of pulses, antenna elements, samples in the fast time domain are $K=100$, $L = 10$ and $N = 100$, respectively. When the 3-D FFT is implemented, data windows are used for both the conventional and one-bit radar to suppress sidelobes of FFT. In the slow time and fast time domain, Chebyshev windows are adopted with their peak sidelobes of $-50$ dB. In the spatial domain, we adopt the Taylor window. The total SNR loss caused by data windows in the three domains can be computed \cite[pp. 257]{Richards} and is about $3.6$ dB.

For one-bit radar, the overgriding factor is $r_a = 4$. As discussed in Section IV, the proposed approach includes two detection processes. Based on the 3-D FFT results, the predetection is firstly  performed to predetect target using the OS CFAR detector at the threshold $\gamma_1$. Then, target detection in the second stage is carried out based on the recovery vector $\hat{{\mathbf x}}_{\kappa}$ of the GAMP algorithm. The detection threshold of the second detector is $\gamma_2$. We generally set a low threshold $\gamma_1$ to implement target predetection. Though the FA rate is high at the predetection stage, many FAs can be removed through the second target detector. Here, the scale factor $\alpha_{OS}$ in ${\gamma}_{1}$  is set as 8.0 dB\footnote{The corresponding FA rate of the predetection is about $6.9\times 10^{-4}$.} and $\eta$ in the spatial, slow time, and fast time domain is set as $18$, $75$ and $75$, respectively. To detect targets based on $\hat{{\mathbf x}}_{\kappa}$ in the second stage, the threshold $\gamma_2$ should be chosen carefully. In this paper, we do not apply the adaptive threshold but a constant threshold for the second detector. That is, $\gamma_2$ is not related to background amplitudes\footnote{After removing the elements corresponding to true targets, the rest of elements in $\hat{{\mathbf x}}_{\kappa}$ are referred as background entries.} of $\hat{{\mathbf x}}_{\kappa}$ and is selected based on the receiver sensitivity. Since the noise power is known, the amplitudes of targets can be recovered using the DR-GAMP approach. Assuming that the minimum target power level expected to be detected is $P_t$, the detection threshold is then given by $\gamma_2 = \sqrt{P_t}$. Here, we set  $\gamma_2 = (G_a-T_h)$ dB\footnote{$\gamma_2$ can be regarded as the receiver sensitivity at $2\sigma_w^2=1$.}, where $G_a = 10{\rm log}(KLN)$ dB denotes the ideal signal processing gain of the receiver and $T_h$ is empirically set as $13.6$ dB.

\begin{figure}[h!t]
\centering
\includegraphics[width=3.5in]{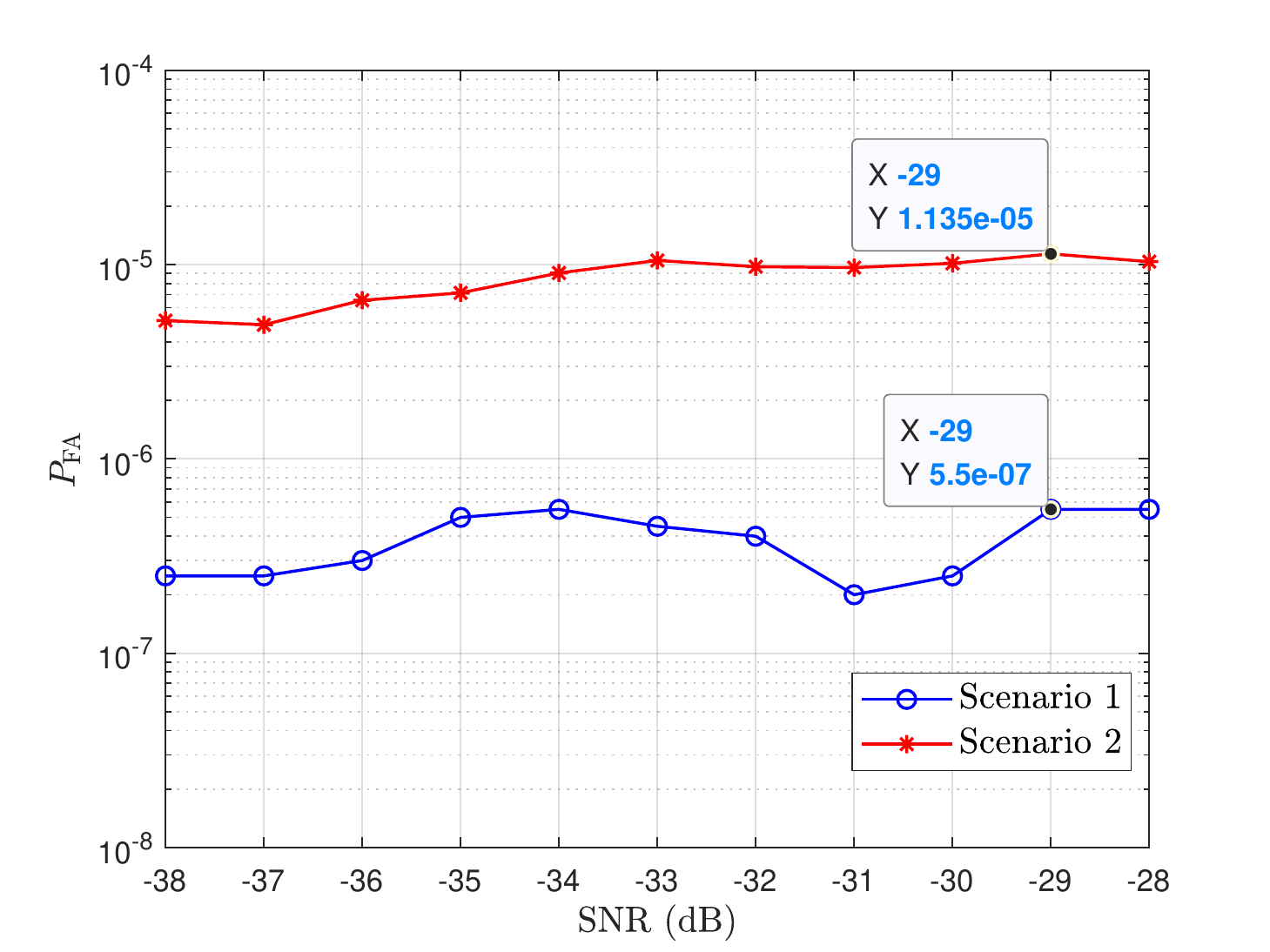}
\caption {FA rates of one-bit radar. (a) Scenario 1. (b) Scenario 2.}
\label{pf_twoScen}
\end{figure}

At given $\gamma_1$ and $\gamma_2$, the FA rates for the two scenarios are shown in Fig. \ref{pf_twoScen}. Results show that the FA rates are related to scenarios\footnote{Note that the predetector in the first stage is CFAR. The detector in the second stage is not CFAR.}. The maximum values for the two scenarios are $5.5\times 10^{-7}$ and $1.1\times 10^{-5}$, respectively.

We carry out performance comparisons of the conventional radar and one-bit radar at $P_{\rm FA,1} = 5.5\times 10^{-7}$ for the scenario 1 and $P_{\rm FA,2} = 1.1\times 10^{-5}$ for the scenario 2, respectively. Based on $P_{\rm FA,1}$ and $P_{\rm FA,2}$, the detection thresholds of the OS CFAR detector for the conventional radar can be obtained through MC simulations. For the scenario 1 and the scenario 2, the scale factor in $\gamma_c$ is $\alpha_{OS,c1} = 11.8$ dB and $\alpha_{OS,c2} = 10.6$ dB, respectively.

\begin{figure}[htb]\centering
\subfigure[]{\begin{minipage}[t]{0.45\linewidth}\centering\includegraphics[width=3.0in]{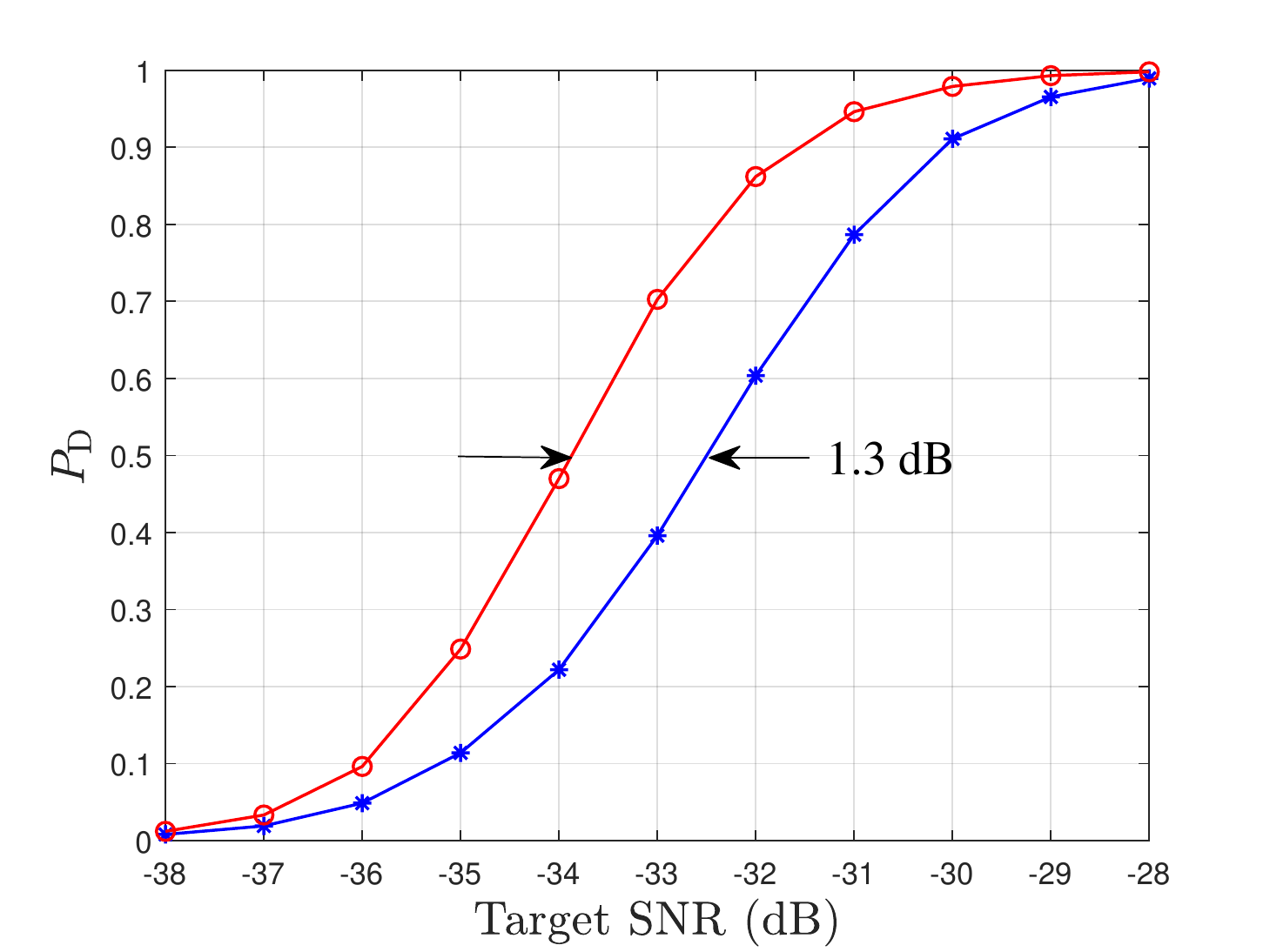} \label{offgridlowSNR0501}
\end{minipage}}
\subfigure[]{\begin{minipage}[t]{0.4\linewidth}\centering\includegraphics[width=3.0in]{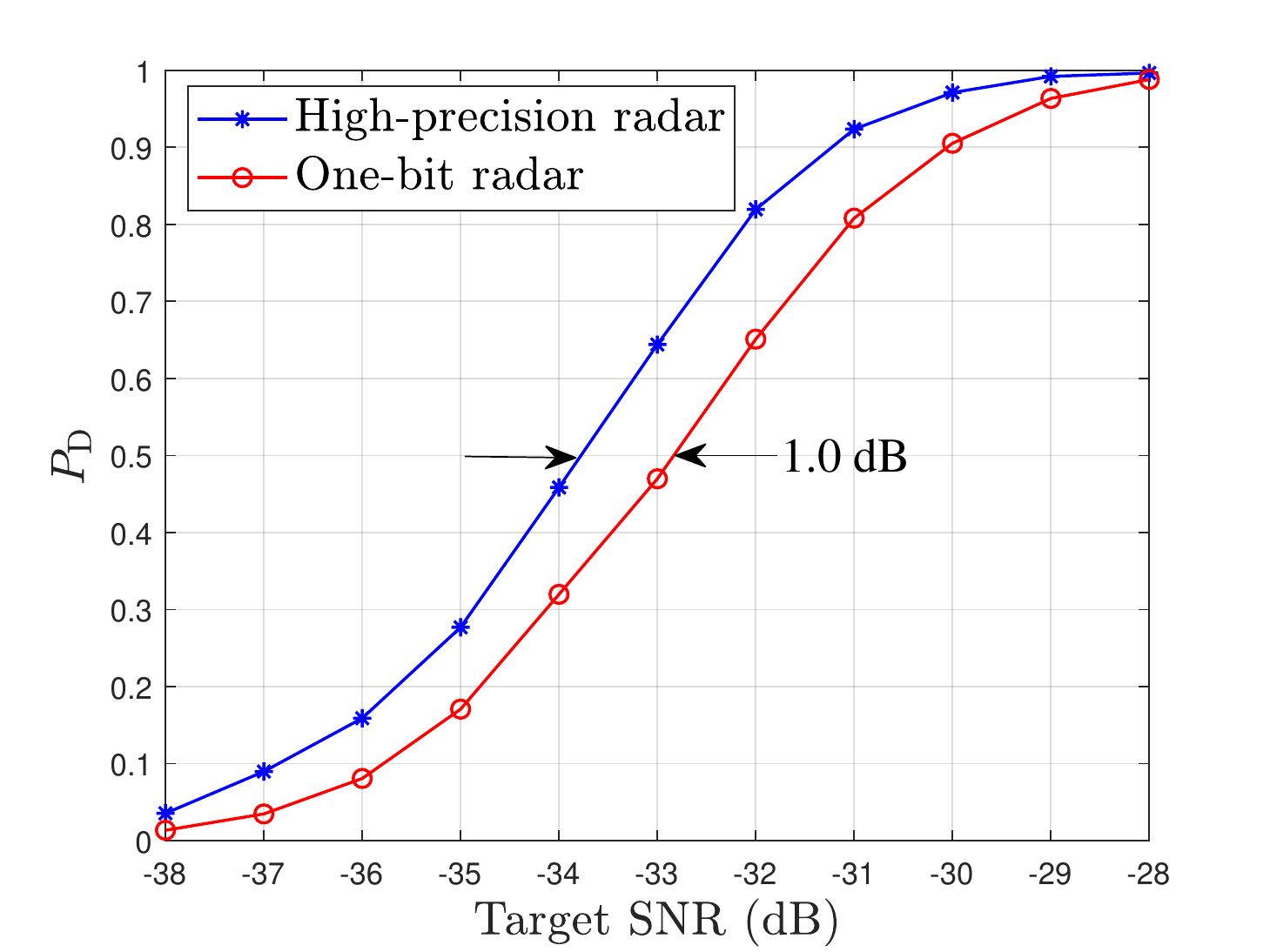}
\label{offgridHighSNR0501}
\end{minipage}}              
\centering {\caption{Performance comparisons.  (a) Scenario 1: Low received SNRs case. (b) Scenario 2: With an additional strong target case. Note that $P_{\rm D}$ is calculated as the detection probability of targets with low SNRs varing from $-38$ dB to $-28$ dB. That is, the strong target is removed when $P_{\rm D}$ is calculated.}}  \label{performanceComparisons}
\end{figure}

The detection performances versus  target SNRs for the both scenarios are presented in Fig. \ref{offgridlowSNR0501} and Fig. \ref{offgridHighSNR0501}. For the scenario 1,  the performance of one-bit radar is superior to the conventional radar, and at $P_{\rm D}=0.5$, the performance gain is about $1.3$ dB. This result seems difficult to understand since one-bit quantization has about 2 dB SNR loss (refer to Fig. 6) at low SNR scenarios compared to the conventional high-precision quantization which exploits more information.
We conjecture that one-bit radar has a lower processing loss compared to the conventional radar.  For the conventional radar, target detection is directly performed based on the 3-D FFT results. The SNR loss of the 3-D FFT  is 3.6 dB (caused by data windows). Nevertheless, one-bit radar has a lower processing loss using the proposed two-stage approach. In the first stage, though the SNR loss of the 3-D FFT is still 3.6 dB, we just perform predetection based on the 3-D FFT results. The predetection threshold $\gamma_1$ is lower than $\gamma_c$ and  this makes the predetection be able to detect targets with lower SNRs than the conventional radar\footnote{For the scenario 1, the scale factor $\alpha_{OS}$ in $\gamma_1$ is 8.0 dB for  one-bit radar. While for the conventional radar, the scale factor $\alpha_{OS,c}$  in $\gamma_c$ is $11.8$ dB. That is, the predetection threshold $\gamma_1$ is 3.8 dB lower than $\gamma_c$. Though one-bit quantization has about 2 dB SNR loss, the predetector in the first stage of the proposed approach can detect targets with  lower SNRs (the value is 3.8 dB-2.0 dB = 1.8 dB) compared with the conventional radar.}. Then, in the second stage, the GAMP algorithm is performed to suppress FAs using the original one-bit observations. There is no weighted loss in the second stage since no data window is used to weight the observations. Therefore, the proposed two-stage approach reduces the processing loss and makes one-bit radar superior to the conventional system for scenario 1.

For scenario 2, the performance of one-bit radar is inferior to the conventional radar, as shown in Fig. \ref{offgridHighSNR0501}.  At $P_{\rm d} = 0.5$, the performance loss is about 1.0 dB under our parameter settings. The reason is that in the presence of a  strong target, the one-bit quantization leads to a larger SNR loss compared to the scenario 1. In the Section V-A-3, Fig. \ref{SNRloss}  indicates that the SNR loss is larger than 2 dB. This larger SNR loss may make the target detection performance of one-bit radar inferior to the conventional radar.

\section{Conclusions}
In this paper, we investigate problems encountered in one-bit LFMCW radar and propose a two-stage DR-GAMP approach for target detection. It is shown that for a  scenario with multiple targets, the 3-order harmonics including self-generated and cross-generated terms can not be omitted. Linear processing methods are ineffective because they can not completely suppress  harmonics especially for 3-order cross-generated terms.  For nonlinear methods, e.g., the GAMP algorithm, are generally impossible to implement because of the large dimension of the observations. As a result, a two-stage DR-GAMP approach combing the linear and nonlinear methods to perform target detection is proposed. In the first stage, the linear preprocessing is carried out to coherently integrate the received data and then, a CFAR detector is performed to predetect targets. Based on the pre-detection results, the number of grid points in the target space is reduced significantly. Then, the GAMP algorithm is proposed to recover true targets. Substantial numerical experiments are conducted to validate the effectiveness of the DR-GAMP approach. Results illustrate that the DR-GAMP approach can suppress high-order harmonics significantly. Besides, it is shown that  under our parameter settings, the performance of one-bit radar is superior to that of the conventional system for low SNR scenarios. At $P_{\rm d}=0.5$, the performance gain is about 1.3 dB. In the presence of a strong target, one-bit radar has a slightly performance loss.

In many applications, clutter are  an important component of radar echoes. When the clutter is taken into account, the coupling among clutter, target and noise are very complex, and it is difficult to obtain analytical solutions because of the highly nonlinearity of the one-bit ADC. Cross-generated harmonics caused by the coupling will result in more complicated influences on the proposed two-stage approach. In the future, a simple clutter scenario, e.g. modeling the clutter as a colored noise, will be firstly considered and then, it is extended to the transmit signal related clutter scenarios. In addition, as shown in Section V-C, the detector in the second stage is not CFAR and the FA rate is related to  scenarios. How to perform target detection based on the DR-GAMP reconstructed results is still an open problem and deserves in-depth study in the future.

\section{Appendix}
\subsection{The average amplitudes of harmonics}\label{SA}
The average value of $v_q(t)$ with respect to $w(t)$ is given by (\ref{decom_ave}). For $m=0$, its average coefficient $c_0$ is
\begin{align}\label{zeroth-order}
c_0=-\frac{\rm j}{\pi}\int_{-\infty}^{\infty}\frac{{\rm exp}\left(-\frac{\sigma_w^2\xi^2}{2}\right)}{\xi}\prod\limits_{p=1}^P   J_0(\xi A_p){\rm d}\xi
\end{align}
For the first kind Bessel function, we have $J_m(x)=J_m(-x)$ for $m$ being even, otherwise $J_m(x)=-J_m(-x)$. As a consequence, we have
\begin{align}\label{zeroth-order-res}
c_0=0.
\end{align}

Similarly, for the self-generated even harmonics, its average amplitudes are zeroes as well. For the odd harmonics, we calculate the case $P=2$. For a general $P$, calculations are very similar but the analytical expressions are hard to obtain.

For $P=2$, two equations \cite[pp. 1096, 6.633, eq.(1)]{tab_int} and \cite[pp. 1097, 6.633, eq.(5)]{tab_int} are utilized.
Without loss of generality, we calculate the coefficient of the $m$-order self-generated harmonics $c_{m,0}\cos(m\omega_1 t+m\Phi_1)$ from the first target $A_1\cos(\omega_1t+\Phi_1)$. The coefficient can be calculated as
\begin{align}\label{m-order-res_appen}
c_{m,0}&=-\frac{2{\rm j}^{m+1}}{\pi}\int_{-\infty}^{\infty}\frac{{\rm exp}\left(-\frac{\sigma_w^2\xi^2}{2}\right)}{\xi}   J_0(A_2\xi)J_{m}(A_{1}\xi ){\rm d}\xi\notag\\
&=-\frac{4{\rm j}^{m+1}}{\pi}\frac{A_1^m(\frac{\sigma_w^2}{2})^{-\frac{m}{2}}}{2^{m+1}\Gamma(m+1)}\sum\limits_{i=0}^{\infty}\frac{\Gamma(i+\frac{m}{2})}{i!\Gamma(i+1)}(-\frac{A_2^2}{2\sigma_w^2})^i\times F\left(-i,-i;m+1;\frac{A_1^2}{A_2^2}\right).
\end{align}
If $A_1=A_2$, (\ref{m-order-res_appen}) can be further simplified as
\begin{align}\label{m-order-res_furth1_appen}
c_{m,0}&=-\frac{2{\rm j}^{m+1}}{\pi}\int_{-\infty}^{\infty}\frac{{\rm exp}\left(-\frac{\sigma_w^2\xi^2}{2}\right)}{\xi}   J_0(A_2\xi)J_{m}(A_{1}\xi ){\rm d}\xi\notag\\
&=-{\rm j}^{m+1}\sqrt{\frac{2}{{\pi}}}\frac{1}{(\frac{m-1}{2})!m}2^{-\frac{3(m-1)}{2}}~_3F_3\left(\frac{m+1}{2},\frac{m}{2}+1,\frac{m}{2};1,m+1,m+1;-\frac{2A_2^2}{\sigma_w^2}\right).\end{align}

For the $m$-order cross-generated harmonic $c_{m_1,m_2}\cos(m_2\omega_2 t+m_2\Phi_2)\cos(m_1\omega_1 t+m_1\Phi_1)$ where $m=m_1+m_2$ ($m_1,m_2\geq 1$), its coefficient can be calculated as
\begin{align}\label{cross-order-res_appen}
&c_{m_1,m_2}=-\frac{4{\rm j}^{m+1}}{\pi}\int_{-\infty}^{\infty}\frac{{\rm exp}\left(-\frac{\sigma_w^2\xi^2}{2}\right)}{\xi}   J_{m_2}(A_2\xi)J_{m_1}(A_{1}\xi ){\rm d}\xi,\notag\\
&=-\frac{8{\rm j}^{m+1}}{\pi}\frac{A_1^{m_1}A_2^{m_2}(\frac{\sigma_w^2}{2})^{-\frac{m}{2}}}{2^{m+1}\Gamma(m_1+1)}\sum\limits_{i=0}^{\infty}\frac{\Gamma(i+\frac{m}{2})}{i!\Gamma(i+m_2+1)}(-\frac{A_2^2}{2\sigma_w^2})^i\times F\left(-i,-m_2-i;m_1+1;\frac{A_1^2}{A_2^2}\right).
\end{align}
When $A_1=A_2$, we have
\begin{align}\label{cross-order-res_furth1_appen}
&c_{m_1,m_2}=-\frac{4{\rm j}^{m+1}}{\pi}\int_{-\infty}^{\infty}\frac{{\rm exp}\left(-\frac{\sigma_w^2\xi^2}{2}\right)}{\xi}   J_{m_2}(A_2\xi)J_{m_1}(A_{1}\xi ){\rm d}\xi\notag\\
=&-{\rm j}^{m+1}\sqrt{\frac{2}{{\pi}}}2^{-m+2}\frac{(m-2)!!}{m_1!m_2!} \left(\frac{A_{1}}{\sigma_w}\right)^m \!\! ~_3  F_3 \!\! \left( \! \frac{m+1}{2},\frac{m}{2}\!+\!1,\frac{m}{2};m_2+1,m_1+1,m+1;-\frac{2A_1^2}{\sigma_w^2}\! \right).
\end{align}

\end{document}